\documentclass[twocolumn,aps,showpacs,nofootinbib,epsfig]{revtex4}
\usepackage{amssymb}
\usepackage{amsmath}
\usepackage{graphics}
\usepackage{epsfig}
\usepackage[dvips]{color}

\newcommand{\be}{\begin{equation}}
\newcommand{\ee}{\end{equation}}
\newcommand{\bea}{\begin{eqnarray}}
\newcommand{\eea}{\end{eqnarray}}
\newcommand{\beas}{\begin{eqnarray*}}
\newcommand{\eeas}{\end{eqnarray*}}

\begin{document}
\title{Monte Carlo approach for hadron azimuthal correlations \\ in
  high energy proton and nuclear collisions} 
\author{Alejandro Ayala$^1$, Isabel Dominguez$^1$, Jamal
  Jalilian-Marian$^2$, J. Magnin$^3$ and Maria Elena
  Tejeda-Yeomans$^4$} \affiliation{$^1$Instituto de Ciencias
  Nucleares, Universidad Nacional Aut\'onoma de M\'exico, Apartado
  Postal 70-543, M\'exico Distrito Federal 04510,
  Mexico.\\ $^2$Department of Natural Sciences, Baruch College, New
  York, New York 10010, USA and CUNY Graduate Center, 365 Fifth
  Avenue, New York, New York 10016, USA.\\ $^3$Centro Brasileiro de
  Pesquisas F\'{\i}sicas, CBPF, Rua Dr. Xavier Sigaud 150, 22290-180,
  Rio de Janeiro, Brazil.\\ $^4$Departamento de F\'{\i}sica,
  Universidad de Sonora, Boulevard Luis Encinas J. y Rosales, Colonia
  Centro, Hermosillo, Sonora 83000, Mexico.}

\begin{abstract}
We use a Monte Carlo approach to study hadron azimuthal angular
correlations in high energy proton-proton and central nucleus-nucleus
collisions at the BNL Relativistic Heavy Ion Collider (RHIC) energies
at mid-rapidity. We build a hadron event generator that incorporates
the production of $2\to 2$ and $2\to 3$ parton processes and their
evolution into hadron states. For nucleus-nucleus collisions we
include the effect of parton energy loss in the Quark-Gluon Plasma
using a modified fragmentation function approach.  In the presence of
the medium, for the case when three partons are produced in the hard
scattering, we analyze the Monte Carlo sample in parton and hadron
momentum bins to reconstruct the angular correlations. We characterize
this sample by the number of partons that are able to hadronize by
fragmentation within the selected bins.  In the nuclear environment
the model allows hadronization by fragmentation only for partons with
momentum above a threshold $p_T^{\mbox{\tiny{thresh}}}=2.4$ GeV.  We
argue that one should treat properly the effect of those partons with
momentum below the threshold, since their interaction with the medium
may lead to showers of low momentum hadrons along the direction of
motion of the original partons as the medium becomes diluted.

\end{abstract}

\maketitle

\section{Introduction}\label{I}

One of the main experimental discoveries in the field of high energy
heavy-ion reactions has been the suppression of particles, with
momentum similar to the leading one, in the away side of azimuthal
angular correlations~\cite{experiments}. It is by now believed that
this phenomenon is caused by the energy loss of partons moving through
the strongly interacting medium formed in the aftermath of the
reaction.  This picture is far from being that simple since when the
difference in momentum between leading and away side particles
increases, either a double peak structure or a broadening of the
away-side peak appears, while either of these are absent in $p + p$
collisions at the same energies~\cite{azcor}.

Given that the structures in the away side are more prominent for
relatively small momentum particles, the above features have generated
different explanations based on medium effects.  The current trend
explains the double peak/broadening in the away side as due to initial
state fluctuations of the matter density in the colliding
nuclei. These fluctuations would in turn be responsible for an
anisotropic flow of partially equilibrated low momentum particles with
the bulk medium.  Recent descriptions of this scenario, based on
hydrodynamics, have reproduced successfully the experimental
$v_3$~\cite{v3}.  Nonetheless, it has also been shown recently that
there is a strong connection between the medium's path length and the
observed away side structures~\cite{pathlength}.  This connection is
expressed through the dependence of the azimuthal correlation on the
trigger particle direction with respect to the event plane in such a
way that, for selected trigger and associated particle momenta, the
double peak is present/absent for out of plane/in plane trigger
particle direction.

The difference in path lengths traveled by partons in the medium is at
the core of the idea that the away side structures may include
contributions of $2\rightarrow 3$ processes. For instance, when the
hard scattering resulting in this kind of processes happens close to
the edge, there is a large chance that at least one of the three final
state partons travels a large distance through the medium therefore,
having a high probability of being absorbed. The resulting hadronic
process has two particles in the final state but its origin is an
underlying partonic event with three partons in the final state, one
of which was unable to hadronize. Conservation of momentum at the
parton level followed by collinear fragmentation gives rise to a
distinctive angular dependence in the azimuthal correlation whereby,
the angular difference between leading and away side particles is
close to $2\pi/3$ radians, regardless of whether one or two partons in
the away side survive their passing through the medium and are able to
hadronize.

The above scenario has been put forward and explored in
Refs.~\cite{Ayala}, using the leading order QCD matrix elements for
$2\rightarrow 3$ parton processes. A serious limitation of this
approach stems from the way the collinear singularities are avoided.
This was implemented by restricting the phase space for parton
production to the regions where the angular difference between leading
and away side partons is far from $0$ and $\pi$. To overcome such
limitation, in this work we present a Monte Carlo approach to study
azimuthal correlations in $p + p$ and $A + A$ collisions.  We use the
MadGraph5~\cite{madgraph} event generator, which includes built-in
functions to cancel collinear and soft divergences for $2\rightarrow
3$ parton processes. To study medium effects we use the modified
fragmentation function approach.  We also study $2\rightarrow 3$
parton processes that contribute to only two hadrons in the final
state.  The work is organized as follows: In Sec.~\ref{II} we present
the basics of the description for three hadron production in $p + p$
and $A + A$ collisions. In Sec.~\ref{III} we introduce a Monte Carlo
event generator to implement the calculation of azimuthal angular
correlations.  To study the details of the effect of energy loss on
partons within different momentum bins, in Sec.~\ref{IV} we present
both the parton and hadron $p_T$ distributions for the Monte Carlo
generated events. We note that for the model parameters used, low
$p_T$ hadrons come from relatively high $p_T$ partons.  We use these
samples to build the azimuthal angular correlations in Sec.~\ref{V}.
Finally, we discuss our results and conclude in Sec.~\ref{VI}, in
particular we argue that the double peak/broadening in the away side
observed in data may still be described within this approach, by
considering the interaction of those partons that are not able to
hadronize outside the medium, but still interact with the bulk
partons.

\section{Three-hadron production}\label{II}

In previous works~\cite{Ayala} we have computed the differential cross
section for three-hadron production in $p + p$ and $A + A$ collisions
at mid-rapidity.  In order to show the essentials of that approach, we
hereby summarize it and refer the reader to the aforementioned
references for further details.

In the case of $p + p$ collisions the differential cross section was
obtained by convoluting over the incoming momentum fraction, the
initial distributions $f_{{\mbox{\tiny{p}}}}$ for partons within the
colliding protons, the matrix elements $|{\mathcal{M}_{2 \rightarrow
    3}}|^2$ and the fragmentation functions
$D_{{\mbox{\tiny{P}}}/{\mbox{\tiny{H}}}}$.  For this we used the CTEQ6
parametrization~\cite{CTEQ6}, the leading order matrix elements
describing the process at the parton level~\cite{Ellis} and KKP
fragmentation functions~\cite{KKP}, for a given total center of mass
energy $\sqrt{s}$ available for the collision.  We considered
collinear fragmentation thus, the angles that define the direction of
the away-side hadrons, $\theta_i^{{\mbox{\tiny{H}}}}$ ($i=2,3$), are
linearly related to the the parton angles $\theta_j$ ($j=2,3$).

To consider the process within a central heavy-ion collision and thus
account for the effects of energy loss, we resorted to the model put
forward in Ref.~\cite{Zhang}. The model considers an initial gluon
density obtained from the overlap of two colliding nuclei, each with a
Woods-Saxon density profile. The gluon density of the medium is
diluted only due to longitudinal expansion of the plasma since
transverse expansion is neglected.

The gluon density $\rho_g$ is related to the nuclear geometry of the
produced medium.  We use the modified fragmentation
functions~\cite{Zhang}
\begin{widetext}
\bea
   \tilde{D}_{{\mbox{\tiny{P}}}_n/{\mbox{\tiny{H}}}_m}(z_{nm})
   =\left(1-e^{-\langle L/\lambda\rangle}\right)
   \left[\frac{z_{nm}'}{z_{nm}}
   D_{{\mbox{\tiny{P}}}_n/{\mbox{\tiny{H}}}_m}(z_{nm}') + 
   \langle L/\lambda\rangle\frac{z_{nm;g}'}{z_{nm}}
   D_{{\mbox{\tiny{P}}}_n/{\mbox{\tiny{H}}}_m}(z_{nm;g}')\right] 
   + e^{-\langle L/\lambda\rangle}
   D_{{\mbox{\tiny{P}}}_n/{\mbox{\tiny{H}}}_m}(z_{nm}),
\label{modfrags}
\eea
\end{widetext}
where $z'_{nm}= h_m/(p_n-\Delta E_n)$ is the rescaled momentum
fraction, of hadron H$_m$ originated from the fragmenting parton
P$_n$, $z_{nm;g}'=\langle L/\lambda\rangle (h_m / \Delta E_n)$ is the
rescaled momentum fraction of the radiated gluon, $\Delta E_n$ is the
average radiative parton energy loss and $\langle L/\lambda\rangle$ is
the average number of scatterings.  Both $\Delta E_n$ and $\langle
L/\lambda\rangle$ are related to the gluon density of the produced
medium. This relation comes in by considering the path length traveled
through the medium by the partons produced in the hard scattering.
The path length is determined from the radial position $r$ and the
angle $\phi$ that the leading parton makes with the radial direction,
within the medium.

The average energy loss $\Delta E_n$ is also proportional to the one
dimensional energy loss $\langle d E_n/d L\rangle_{1d} $ parametrized
as 
\bea \Bigl\langle \frac{d E_n}{d L}\Bigr\rangle_{1d} &=& \epsilon_0
\Bigl[\frac{p_n}{\mu_0} - 1.6\Bigr]^{1.2}\Bigl[7.5 +
  \frac{p_n}{\mu_0}\Bigr]^{- 1}.
\label{dEdL}
\eea 
The one-dimensional energy loss per unit length parameter,
$\epsilon_0$, is related to the mean free path $\lambda_0$ by
$\epsilon_0\lambda_0=0.5$ GeV.  Notice that Eq. (2) means that there
is a lower threshold for $p_n$ given by $p_T^{\mbox{\tiny{thresh}}} =
1.6 \times \mu_0 = 2.4$ GeV below which the one-dimensional energy
loss becomes meaningless.  This is interpreted as reflecting the fact
that the energy loss model cannot produce hadrons by fragmentation for
partons below this momentum threshold.  Those partons should thus be
absorbed by the medium and their effect has to be accounted for by
other hadronization mechanisms.

The above described procedure has an inherent limitation which we have
already identified in Refs. \cite{Ayala}.  These are related to the
angular cuts required to avoid the collinear singularities when two of
the final state hadrons are either parallel or antiparallel. In order
to overcome such limitations we proceed to present a Monte Carlo
implementation of two and three hadron production in $p + p$ and $A +
A$ environments dedicated to the study of azimuthal angular
correlations.

\section{Monte Carlo event generator with energy loss}\label{III}

\subsection{Parton level event generator}

We implement the production of parton events using
MadGraph5~\cite{madgraph}, in $p + p$ collisions at RHIC energies at
mid-rapidity ($|\eta|<0.5$).  MadGraph5 generates the helicity
amplitude subroutines to build the hard scattering matrix elements,
used as inputs for the partonic event generator.  It also generates a
set of functions, inspired by the dipole subtraction
method~\cite{cataniseymoursoper}, that remove the soft and collinear
singularities in the phase-space integrated matrix elements arising
under certain kinematical conditions, producing an unweighted set of
hard scattering partonic events. A typical run involves declaring the
values for certain parameters such as the beam energy, the
factorization scale $\mu^2$, the fragmentation scale $Q^2$, the $p_T$ bin
size of the final state partons and the value of parameters like {\tt
  xqcut} that controls the numerical cancellation of the
singularities, according to the particular kinematics. It is worth
mentioning that, as suggested in the literature \cite{madgraph}, we
use values for {\tt xqcut} below 10 GeV.

\begin{figure*} 
{\centering
  {\includegraphics[scale=0.3]{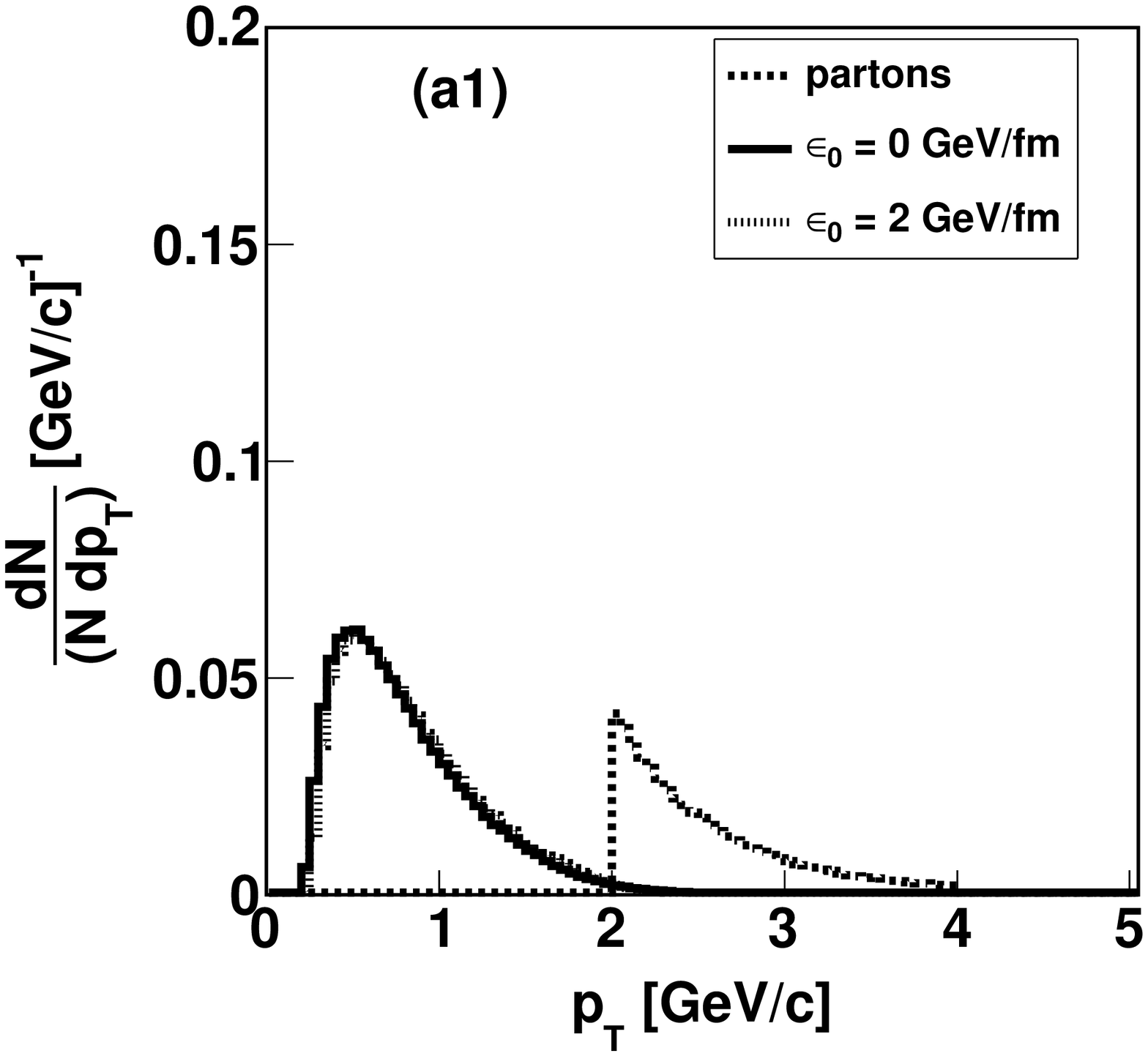}\includegraphics[scale=0.3]{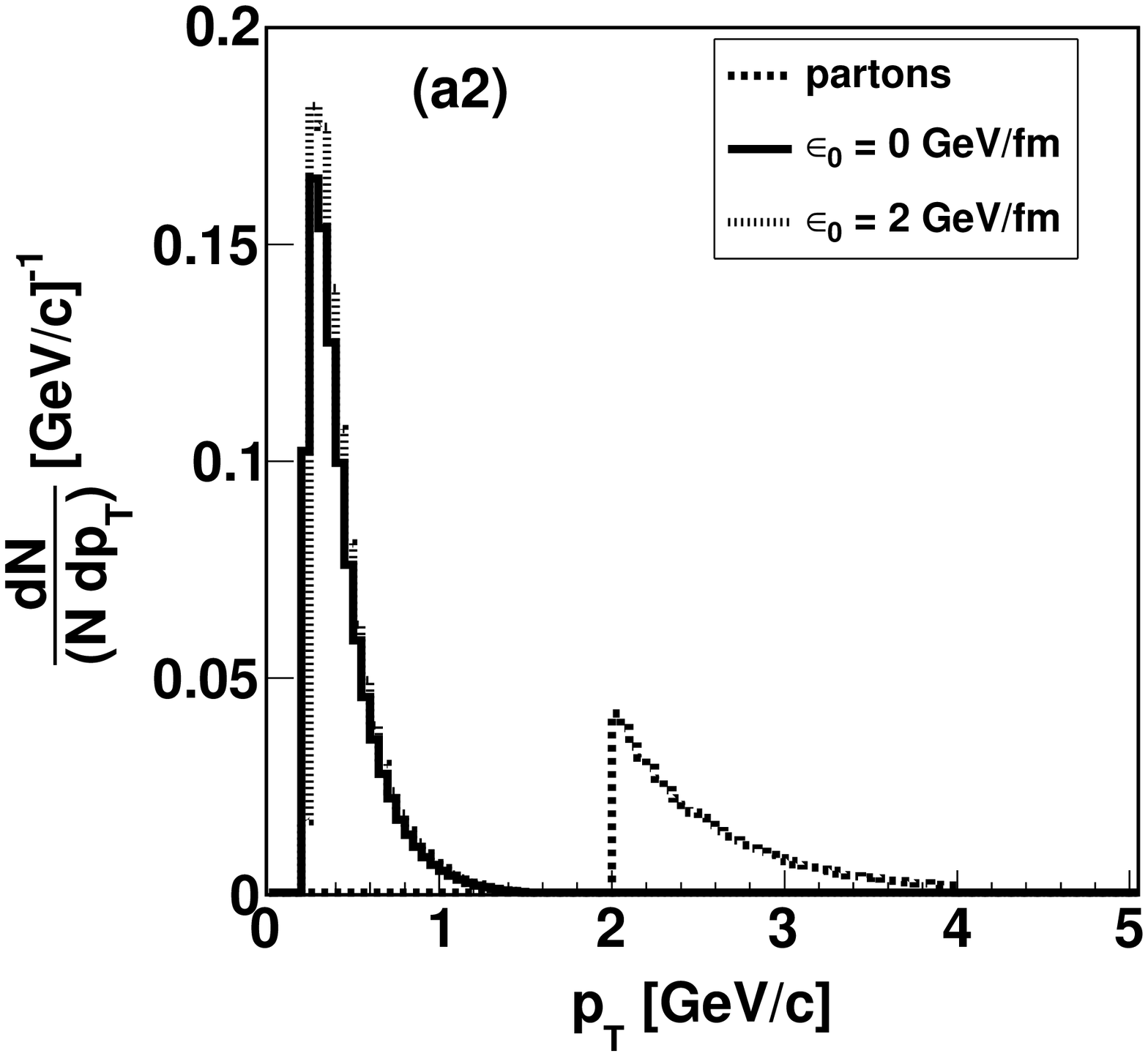}
    \includegraphics[scale=0.3]{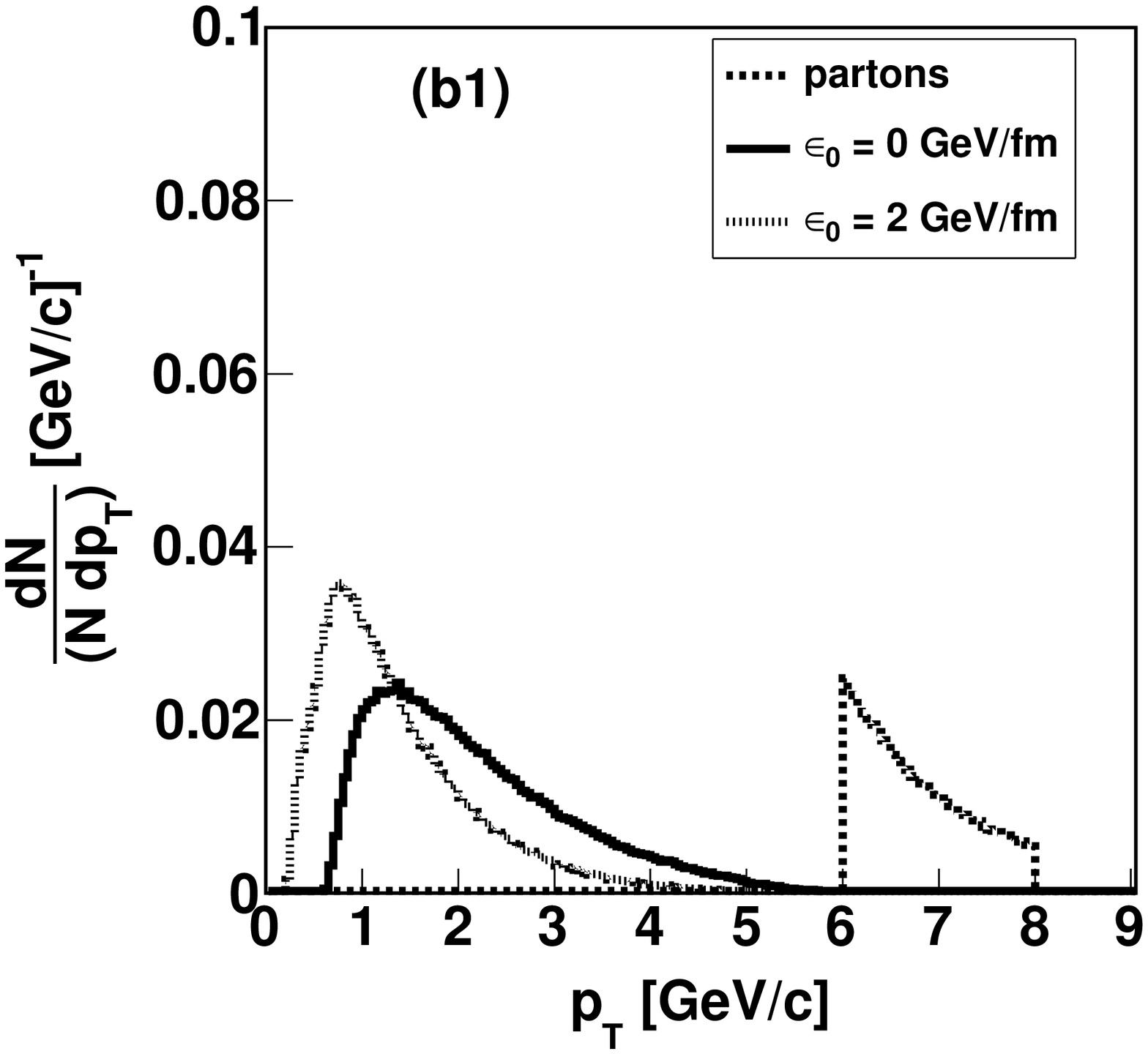}\includegraphics[scale=0.3]{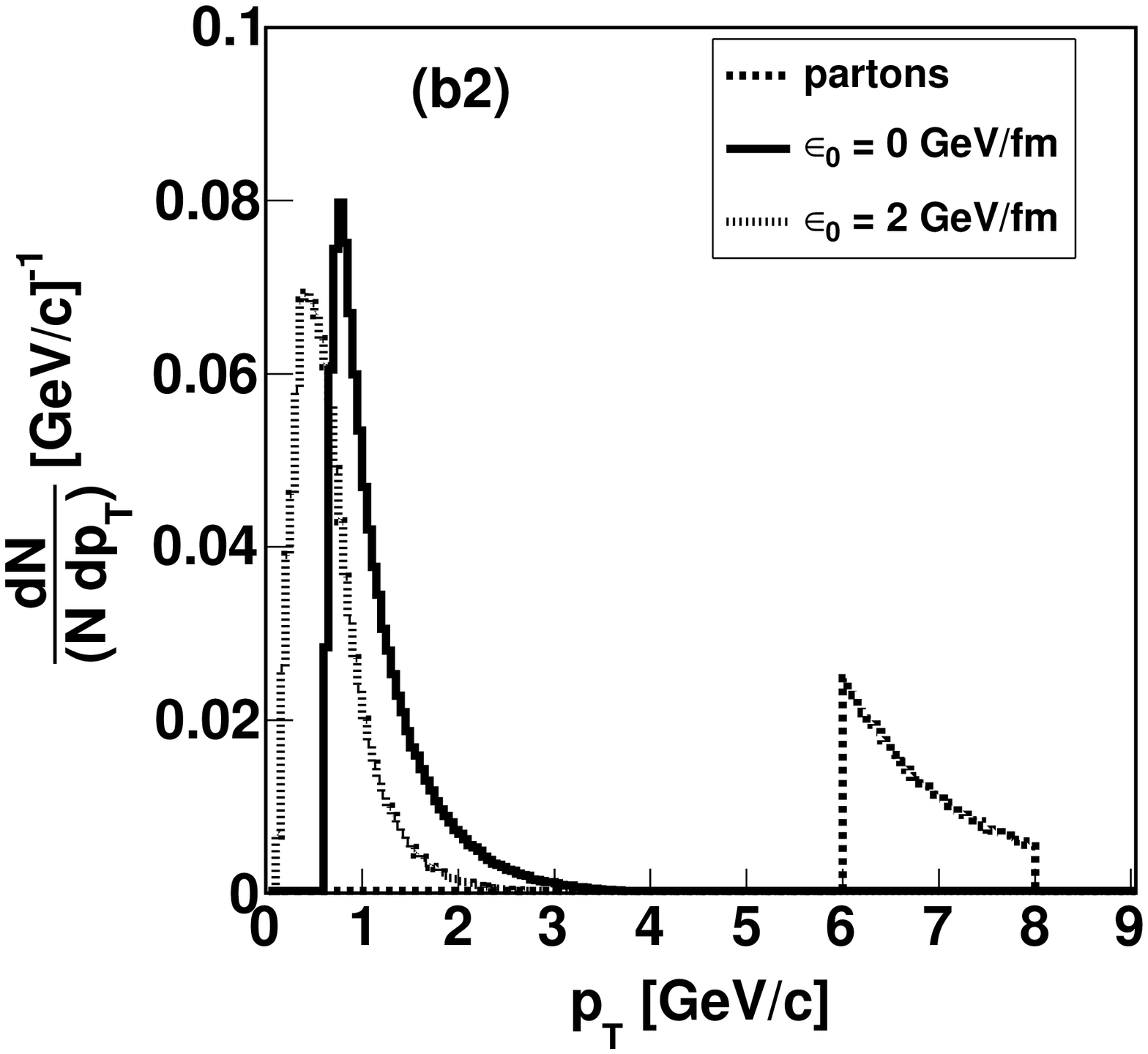}
    \includegraphics[scale=0.3]{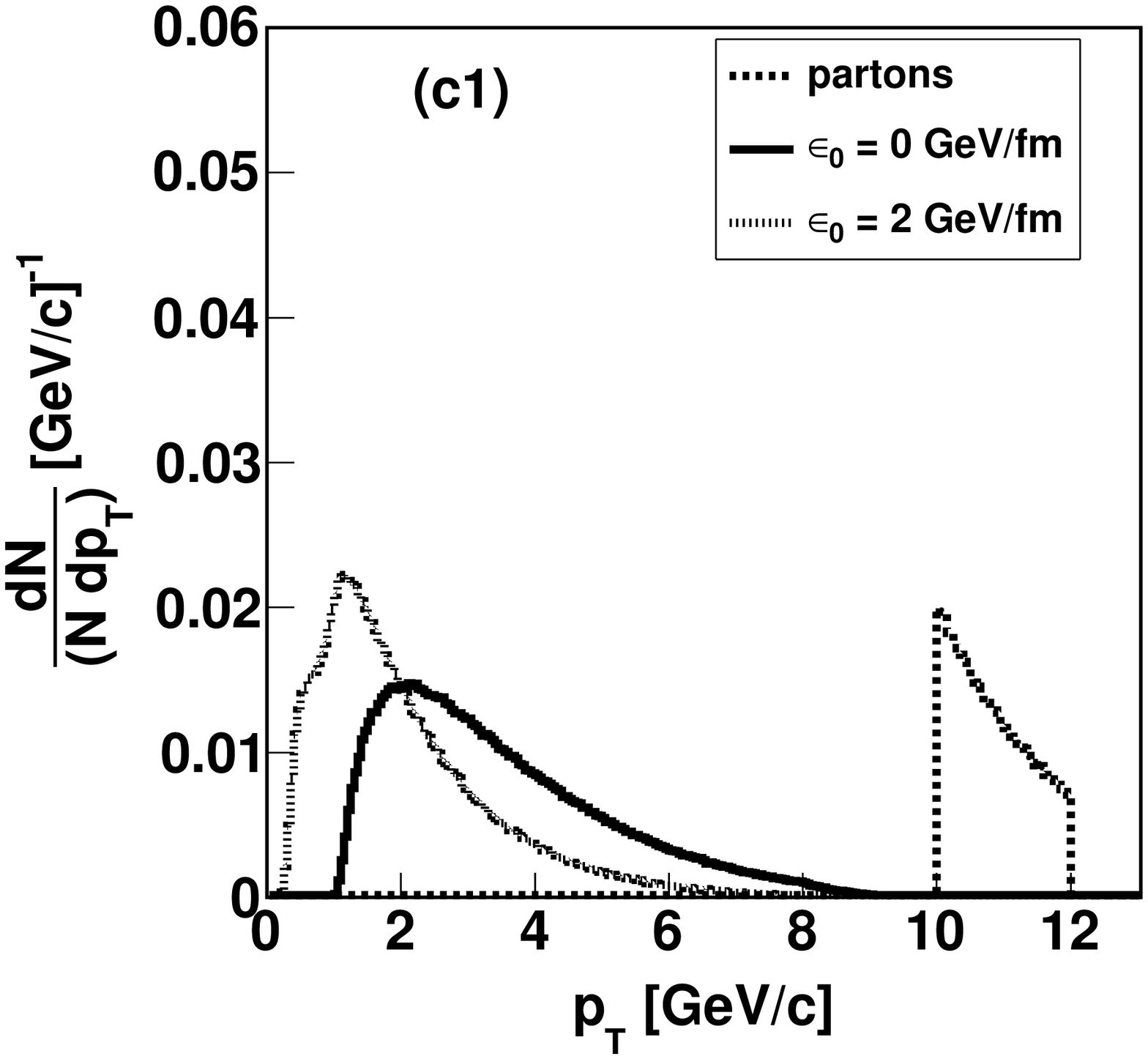}\includegraphics[scale=0.3]{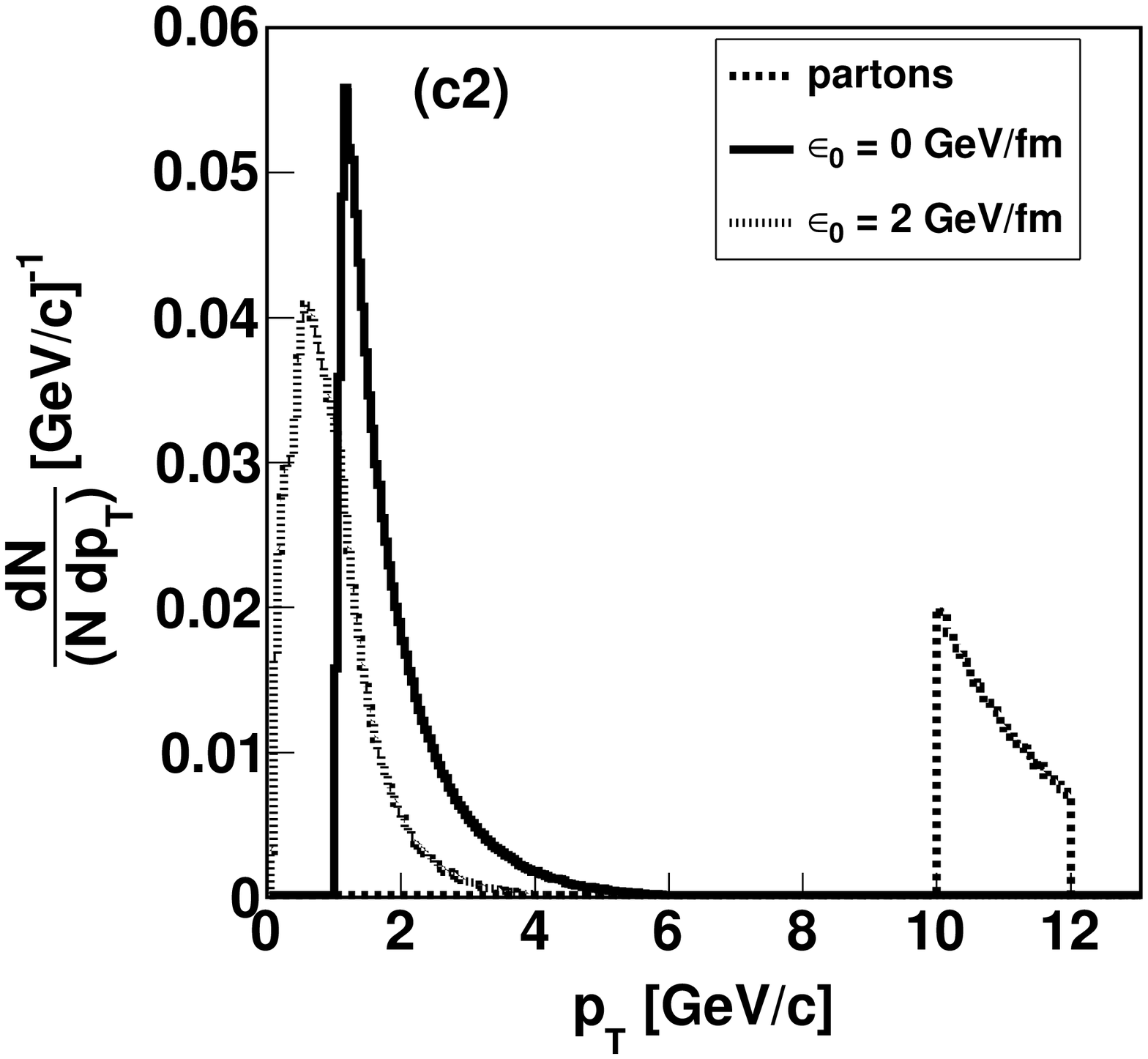}
    \includegraphics[scale=0.3]{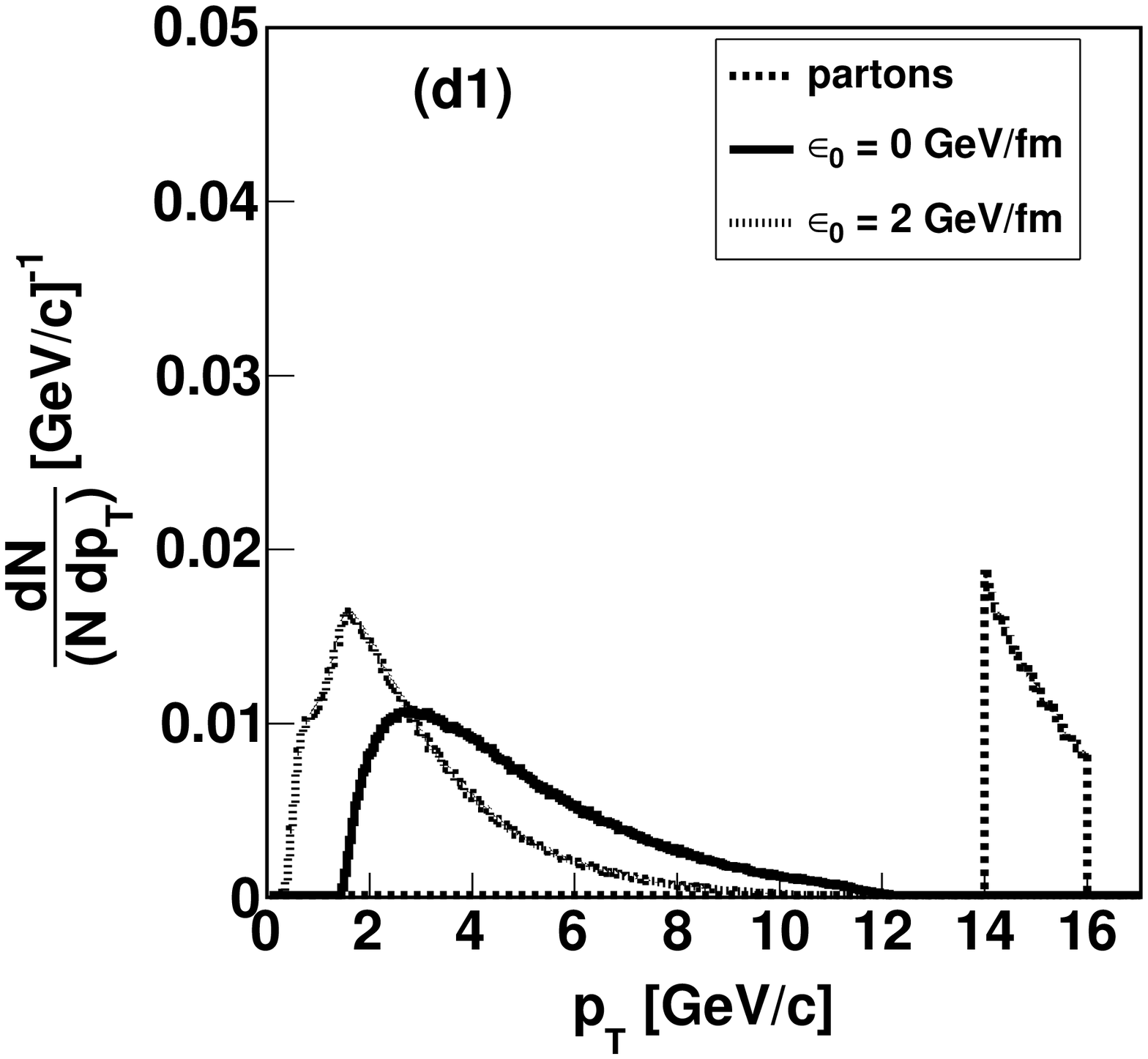}\includegraphics[scale=0.3]{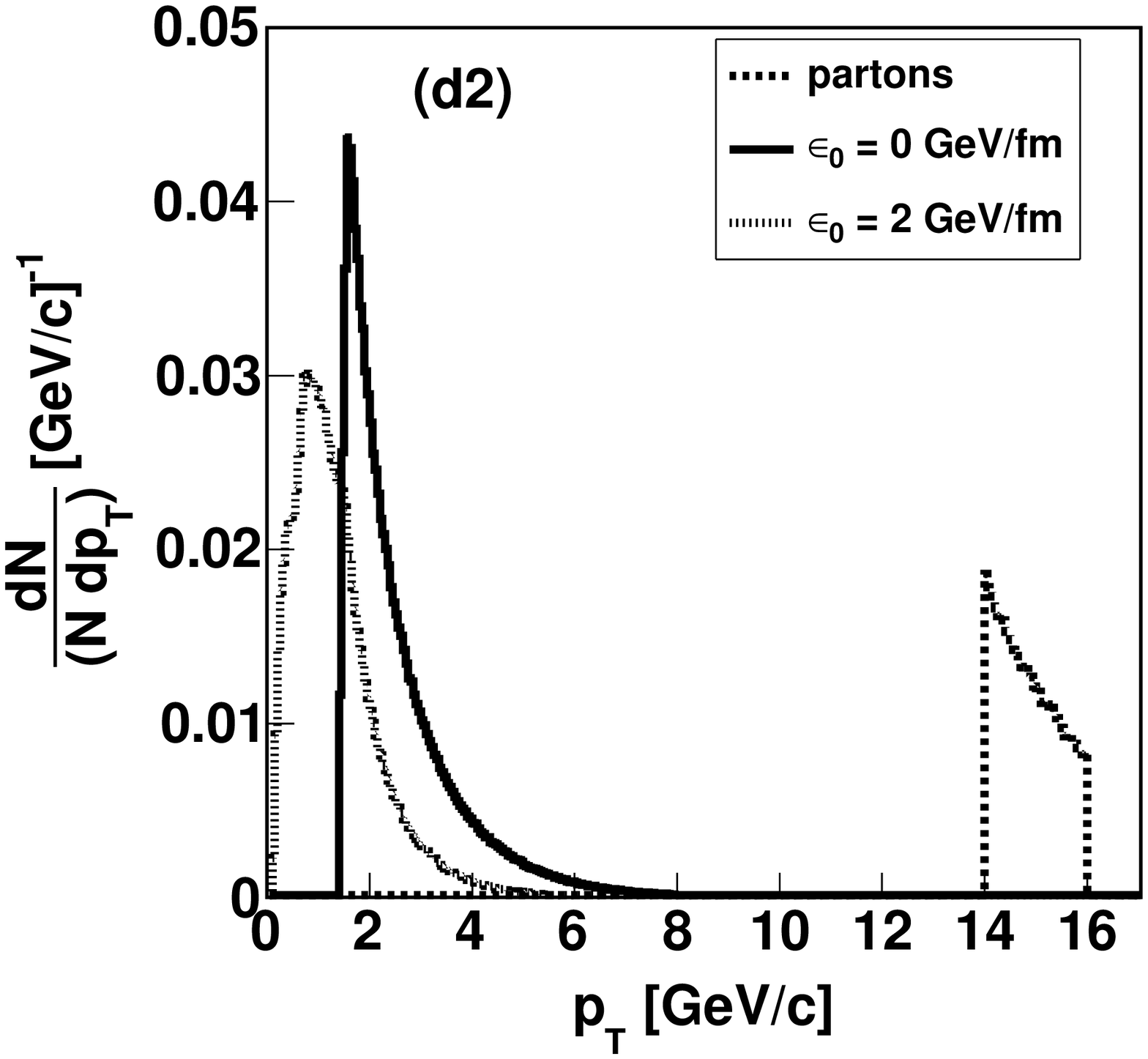}
}}
\caption{Parton (dotted lines) and hadron $p_T$ distributions for $2
  \to 2$ events in $p + p$ ($\epsilon_0=0$, solid lines) and in $A +
  A$ ($\epsilon_0=2\ \mbox{GeV}/\mbox{fm}$, dash-dotted lines) collisions for
  central rapidity $|\eta| \leq0.5$ at $\sqrt{s_{NN}}=200$ GeV.  On
  the 1 (2) column we show the $p_T$ distribution for the
  leading (away) hadron. The hadrons come from partons produced in four
  different momentum bins, from top to bottom: $2-4$ (a), $6-8$ (b),
  $10-12$ (c) and $14-16$ (d) $\mbox{GeV}$. }
\label{fig1}
\end{figure*}

\subsection{Hadron level event generator}

In order to achieve hadron events we take the parton level events and
evolve the partons into hadrons with collinear fragmentation.

\begin{figure*} 
{\centering
  {\includegraphics[scale=0.28]{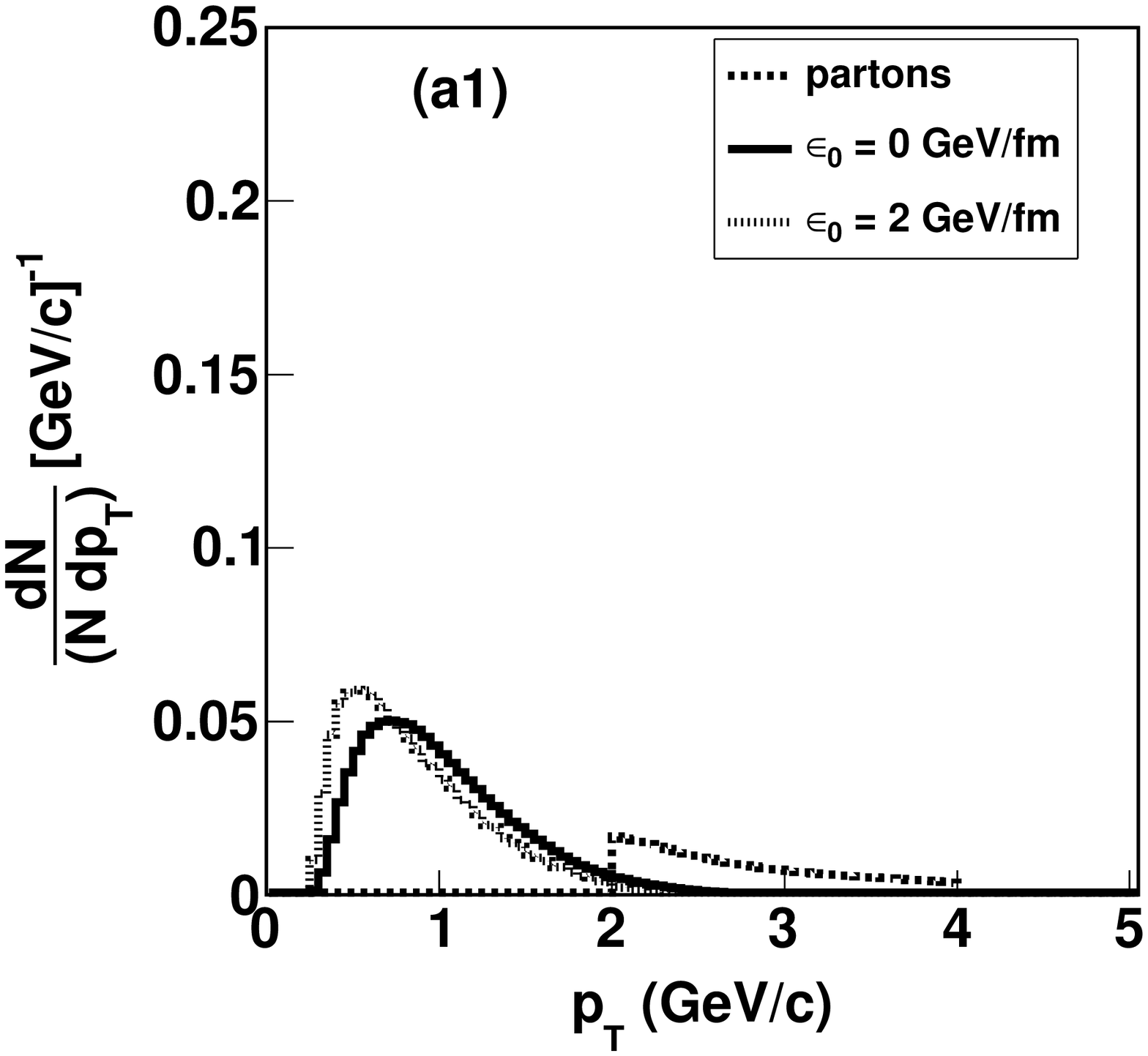}\includegraphics[scale=0.28]{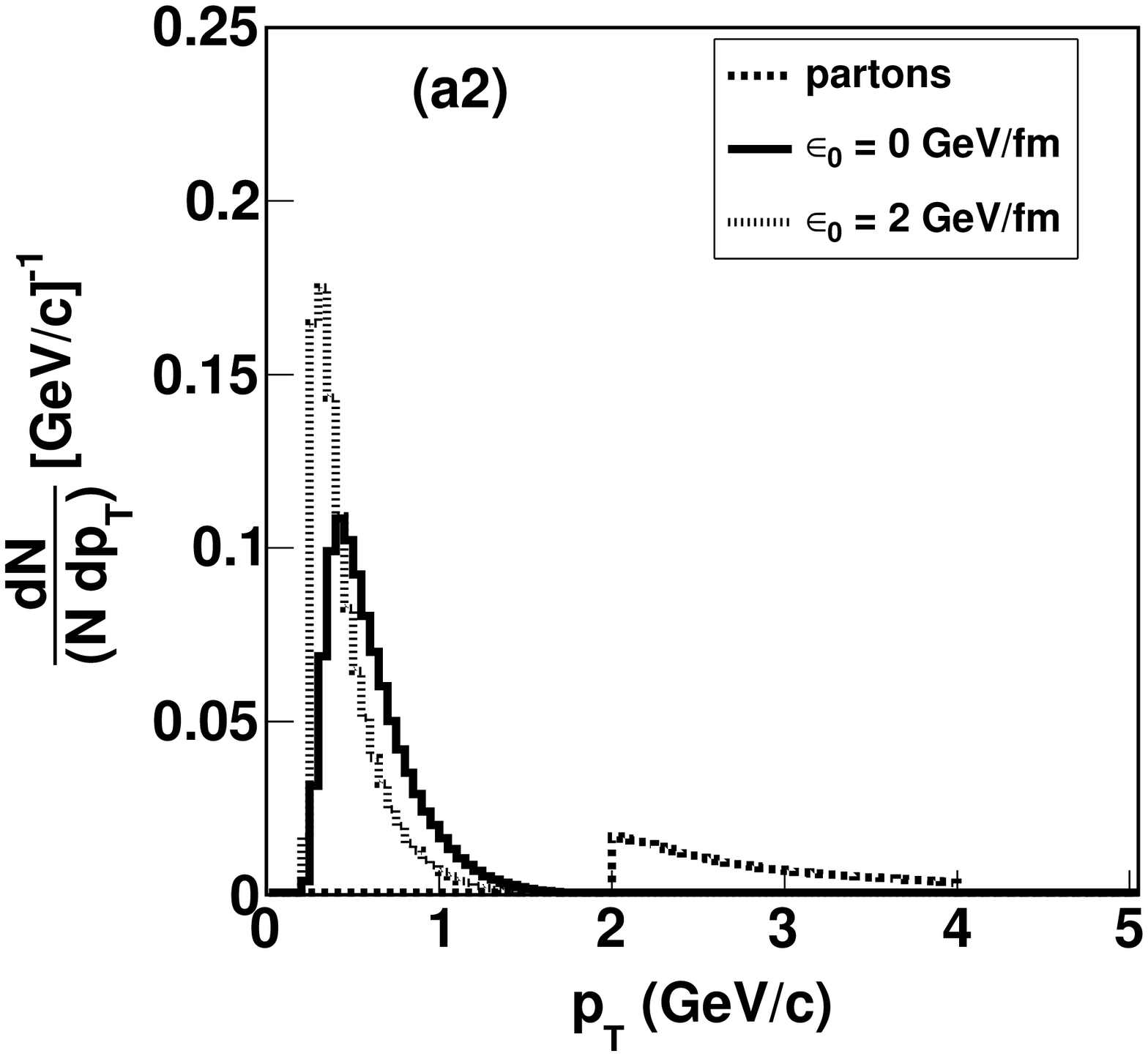}\includegraphics[scale=0.28]{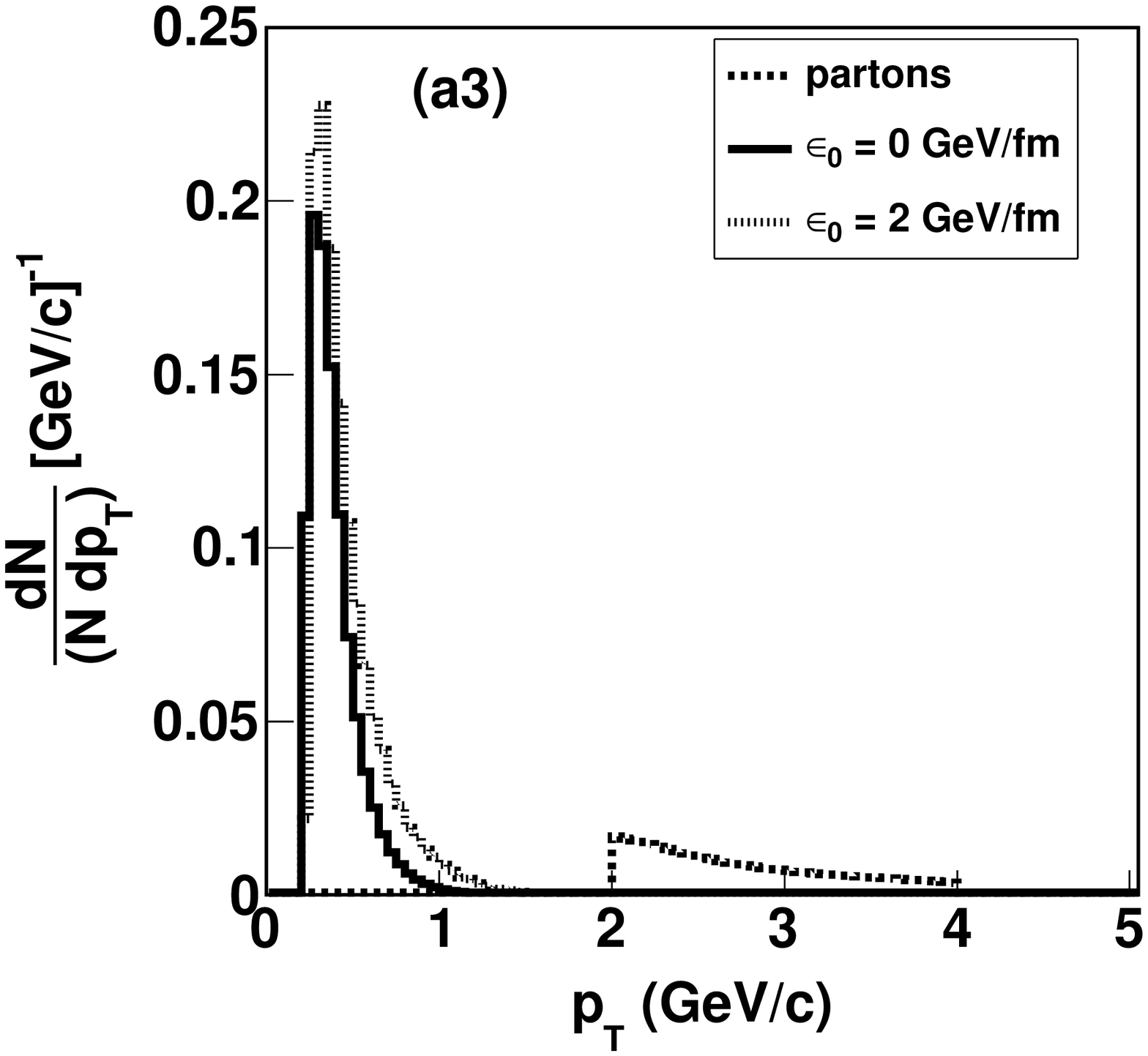}
    \includegraphics[scale=0.28]{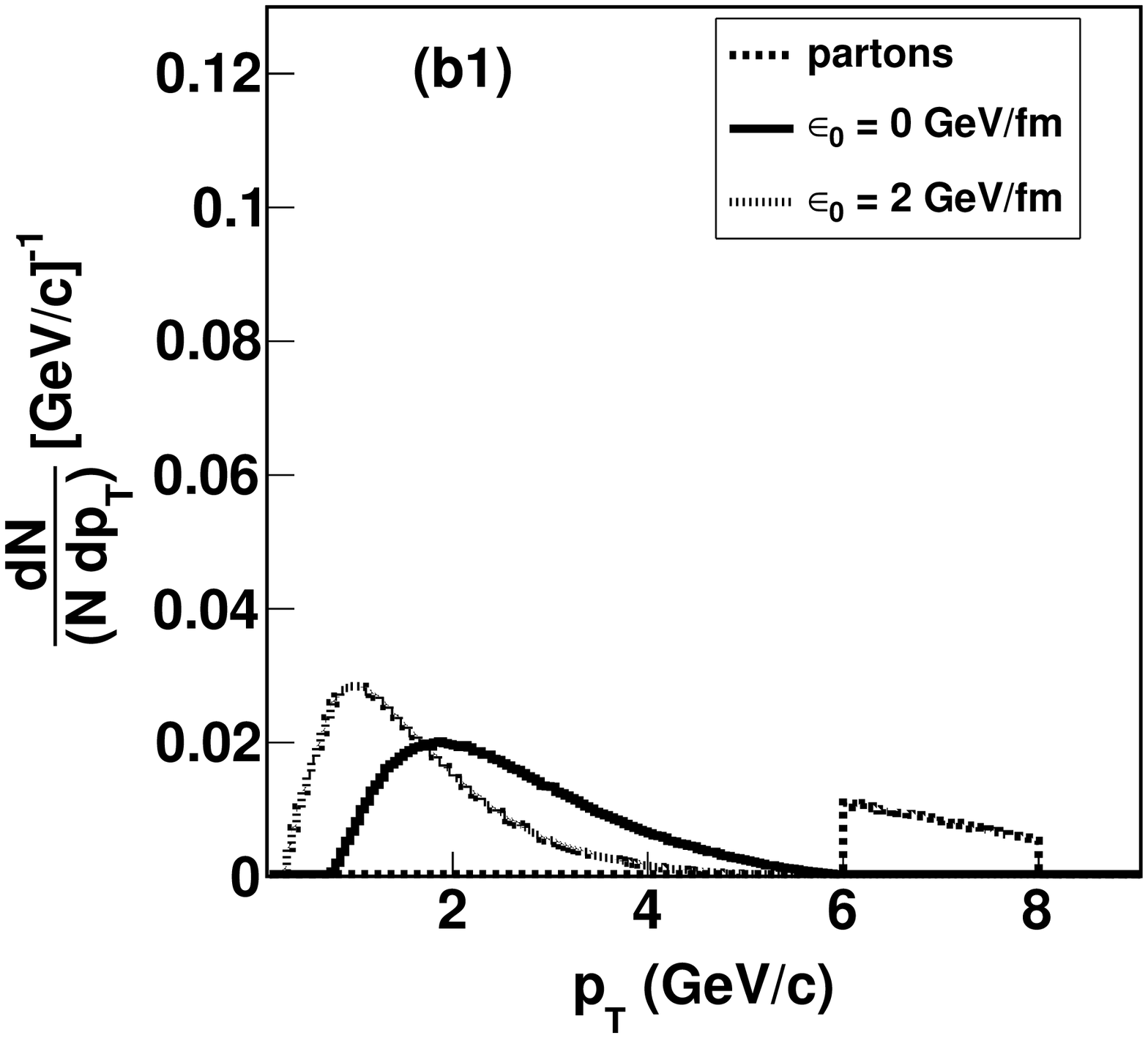}\includegraphics[scale=0.28]{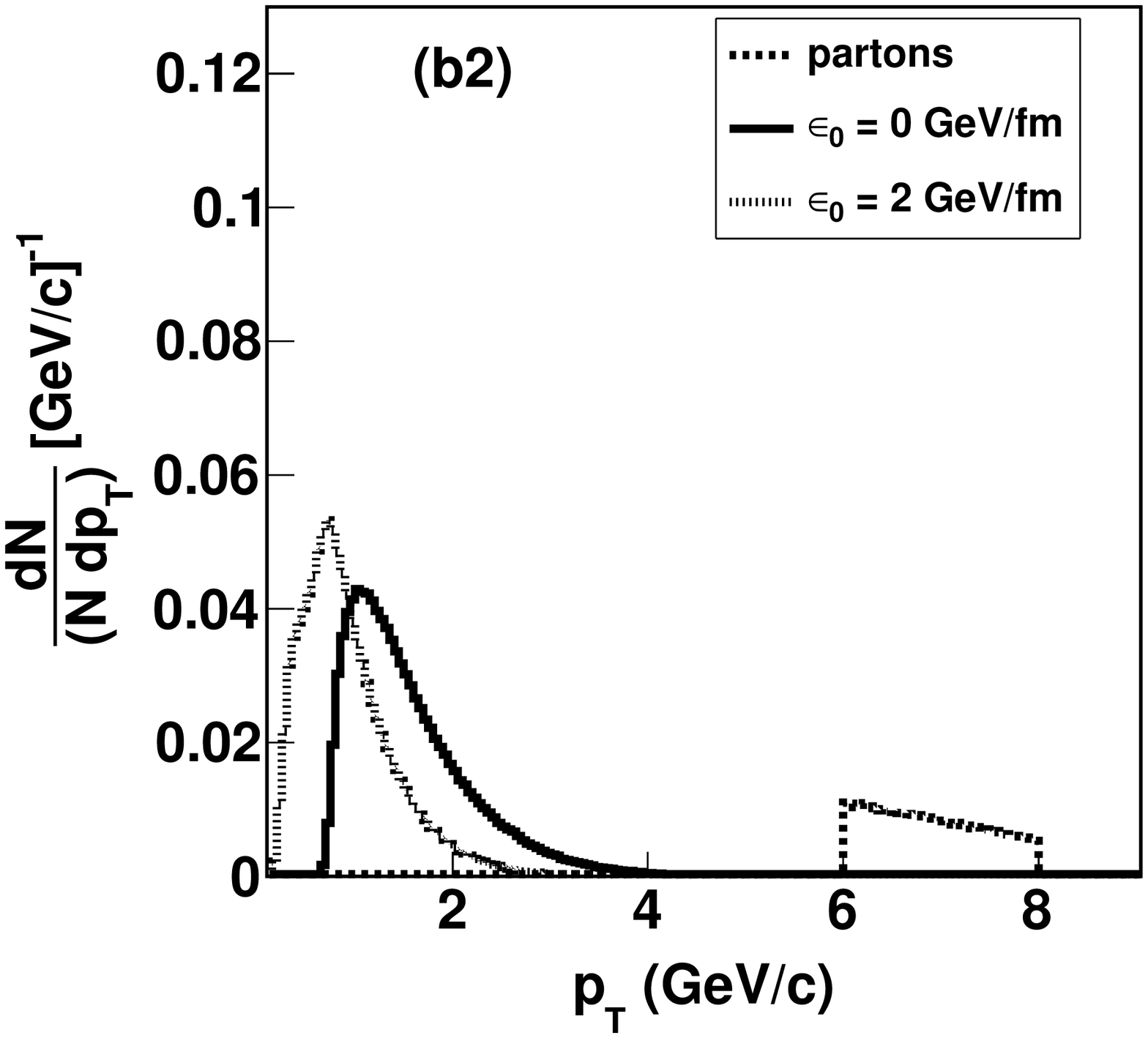}\includegraphics[scale=0.28]{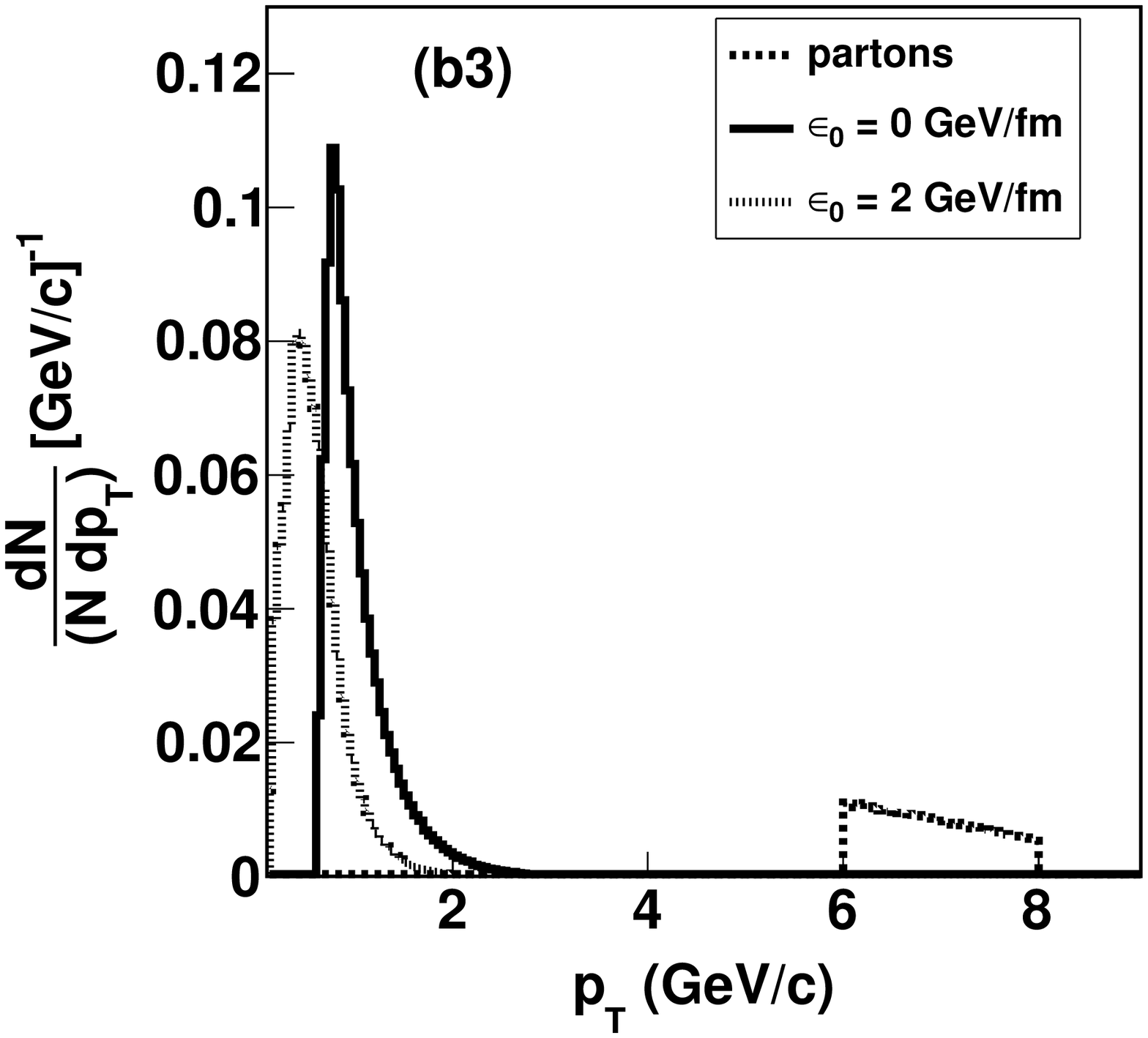}
    \includegraphics[scale=0.28]{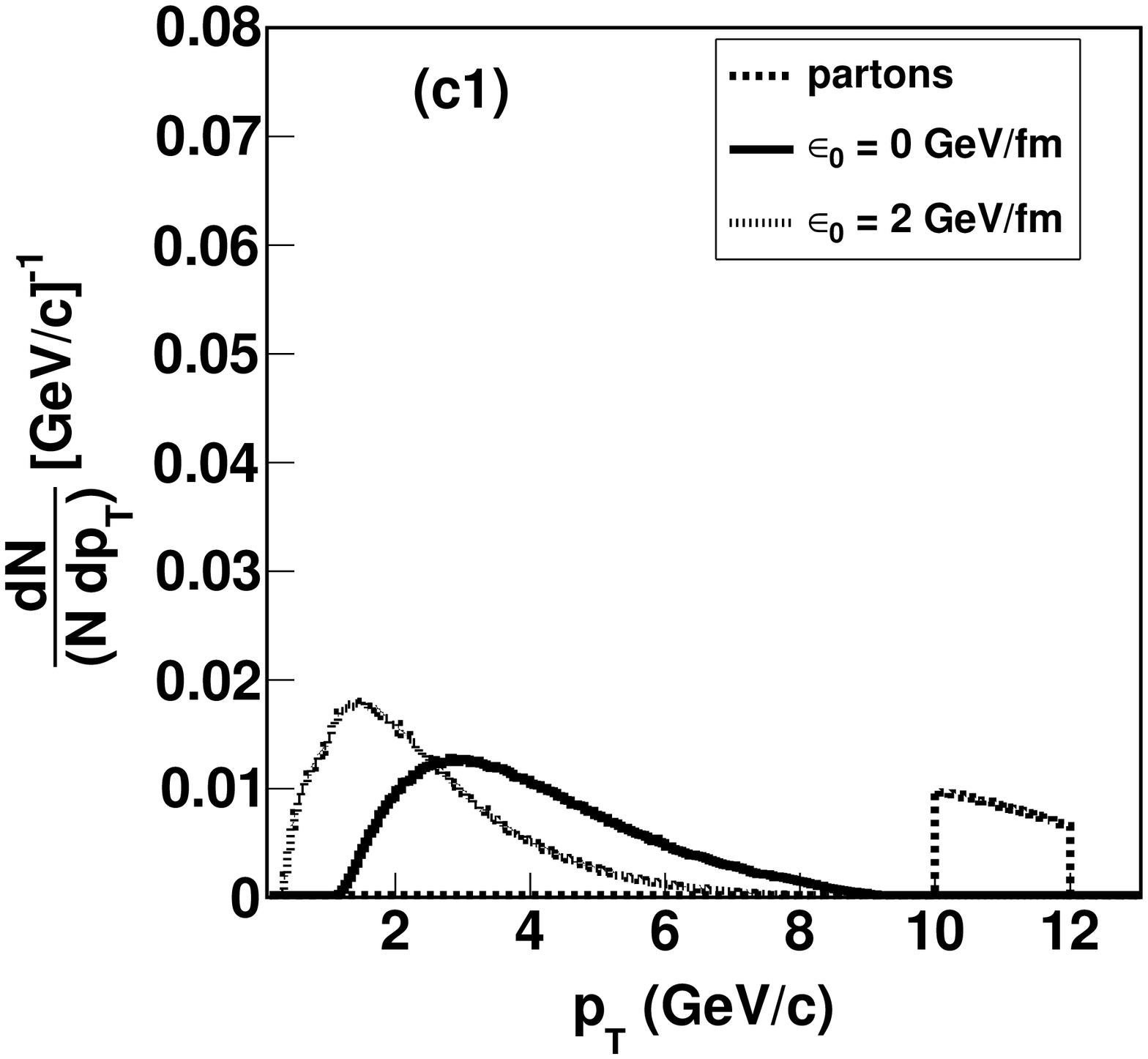}\includegraphics[scale=0.28]{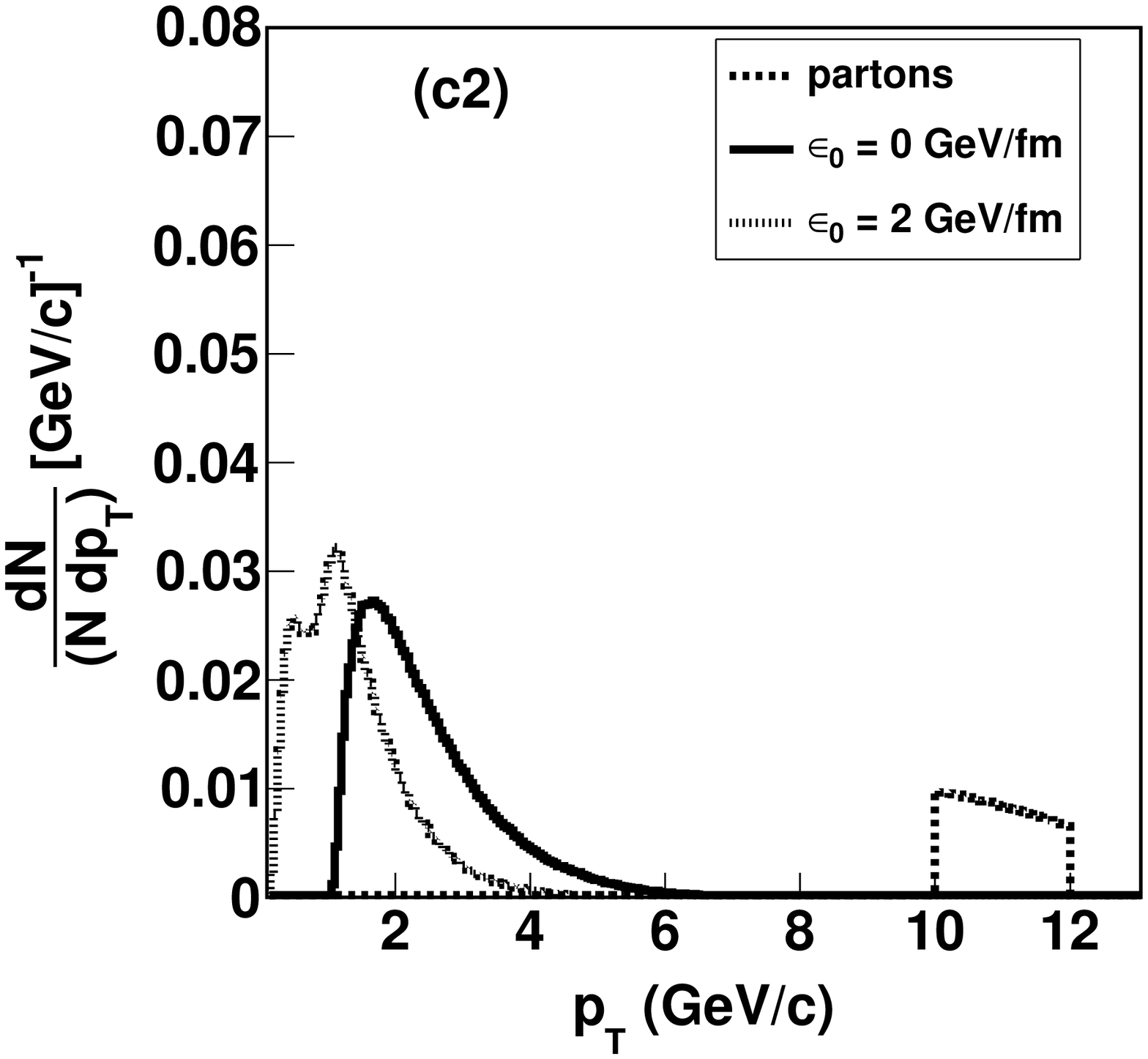}\includegraphics[scale=0.28]{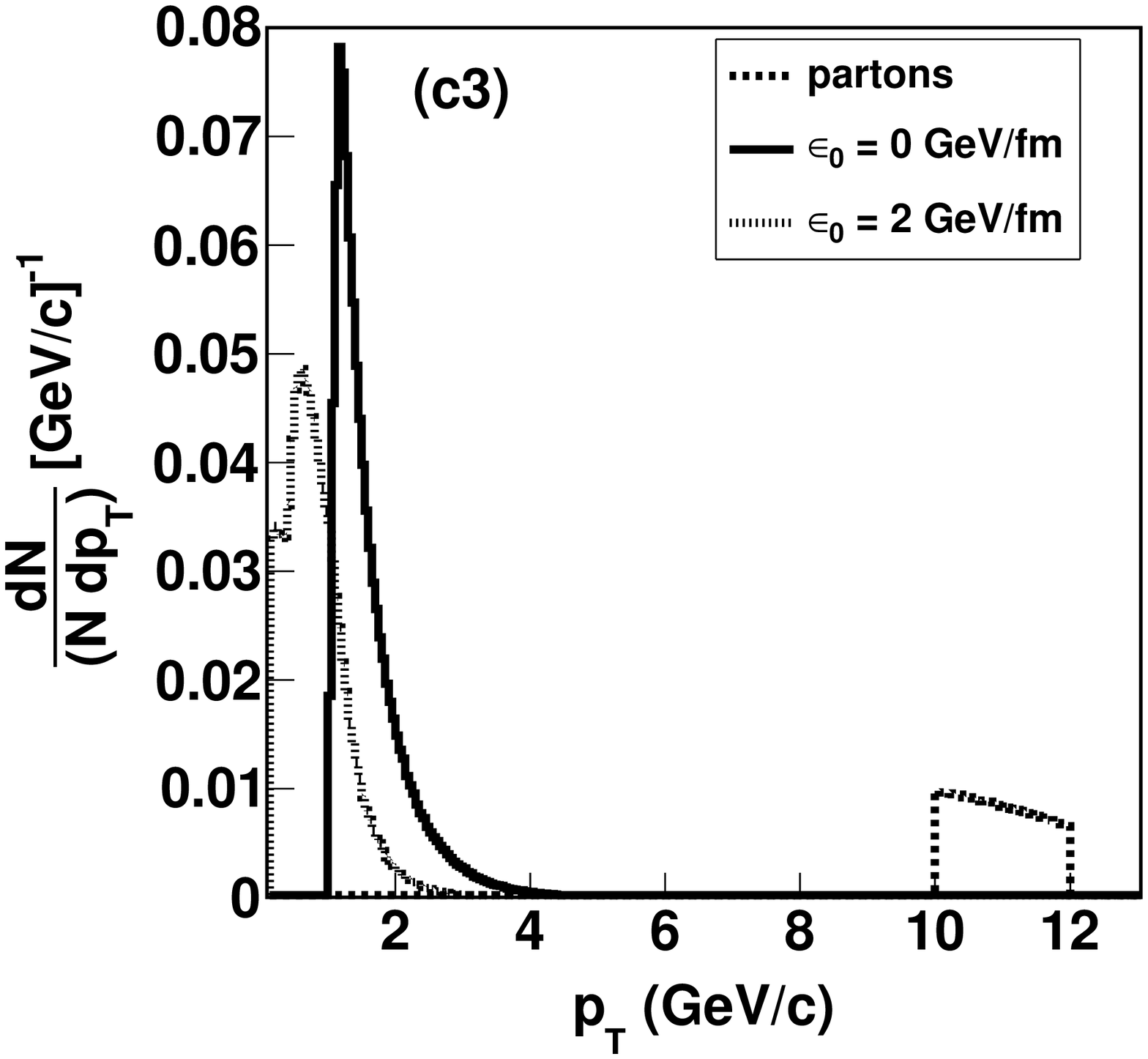}
    \includegraphics[scale=0.28]{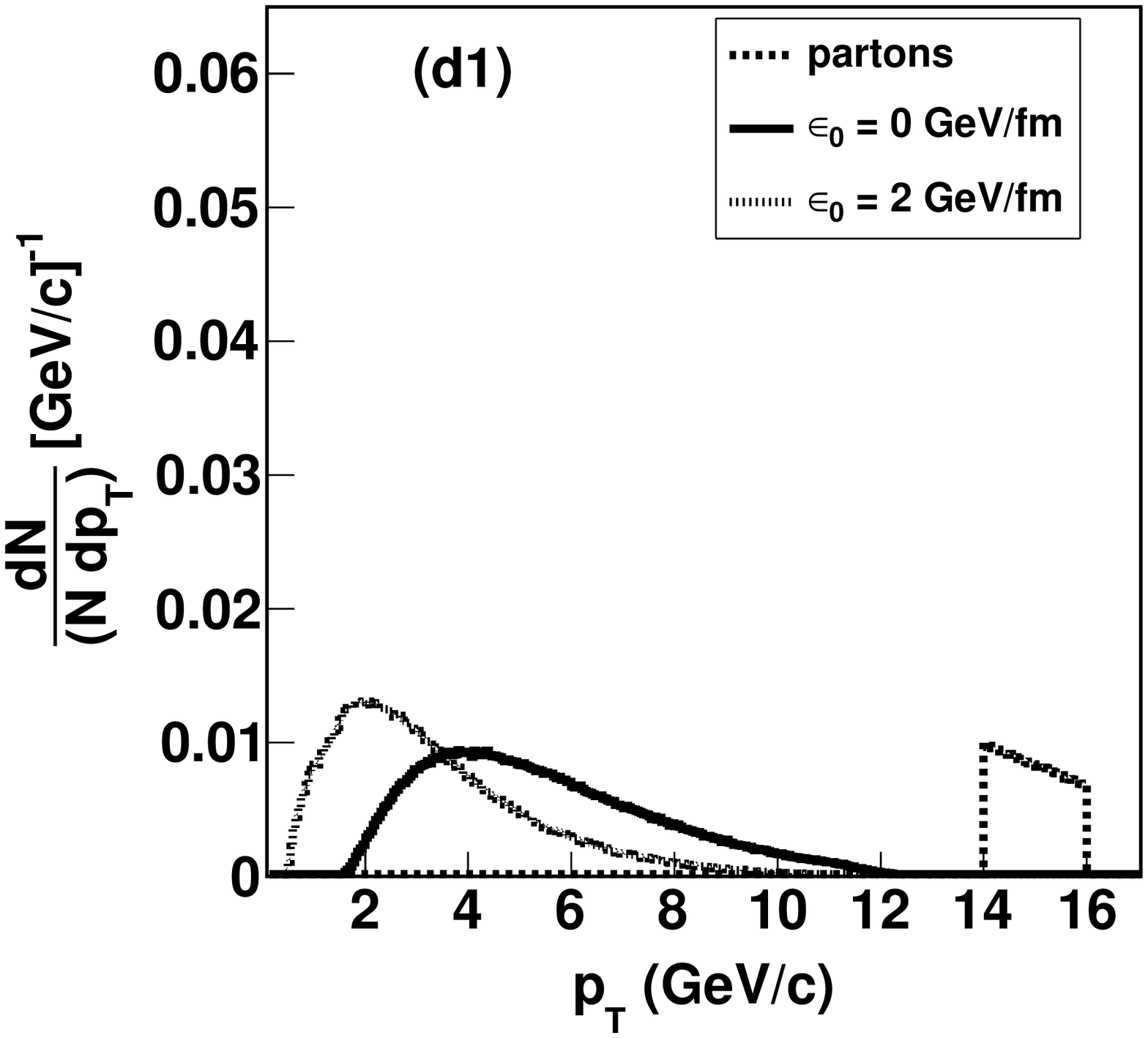}\includegraphics[scale=0.28]{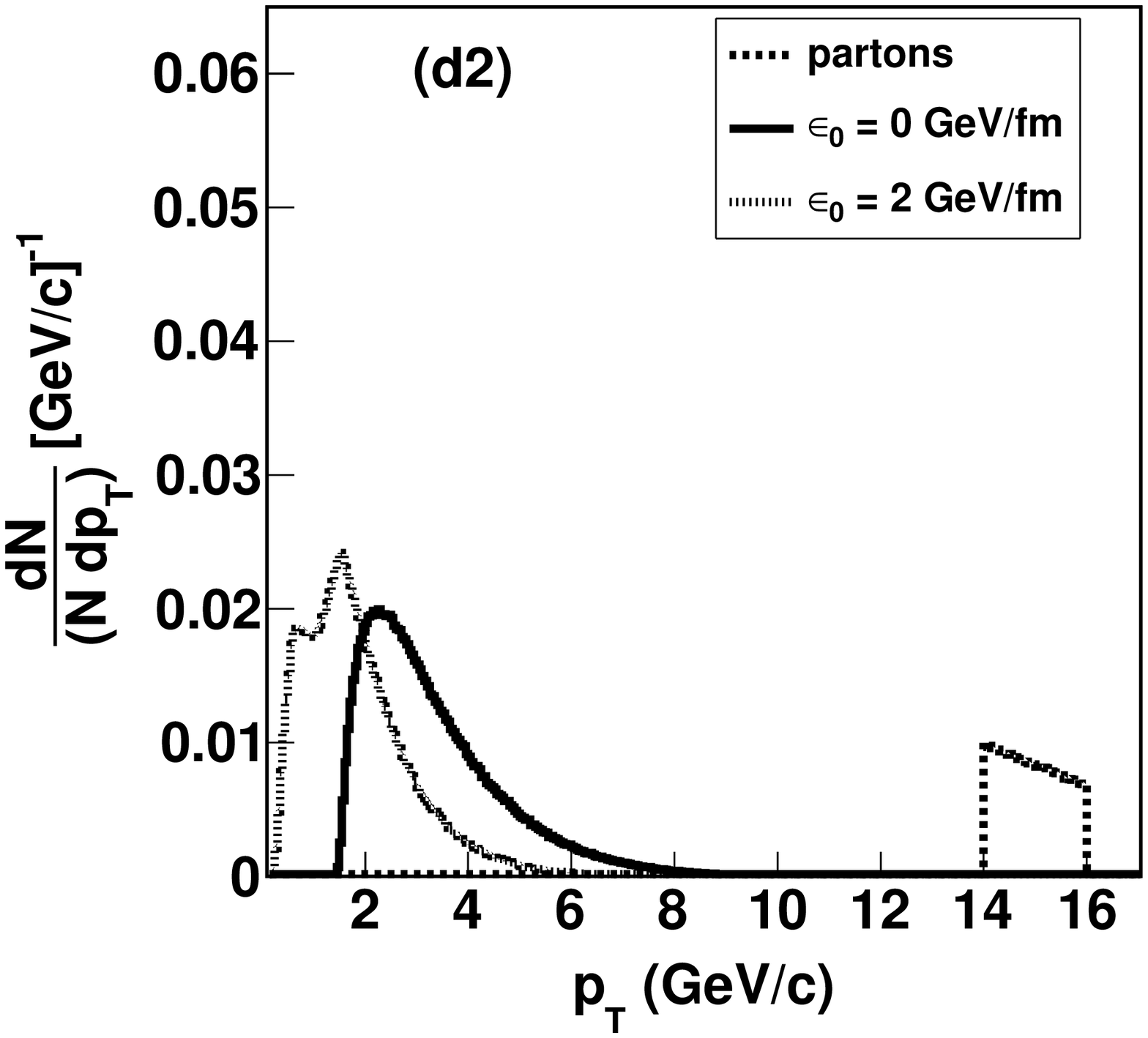}\includegraphics[scale=0.28]{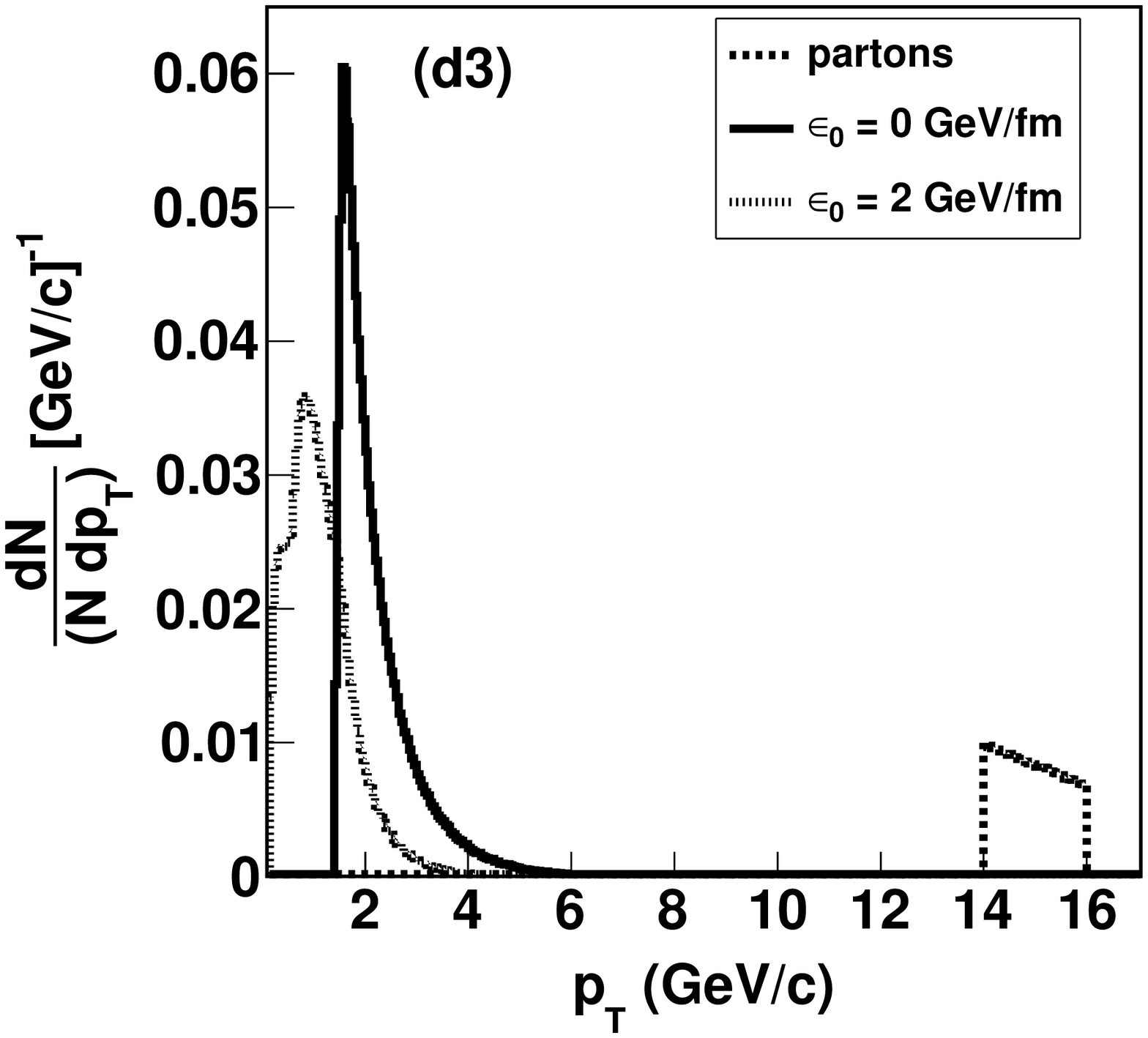}
}}
\caption{Parton (dotted lines) and hadron $p_T$ distributions for $2
  \to 3$ events in $p + p$ ($\epsilon_0=0$, solid lines) and in $A +
  A$ ($\epsilon_0=2\ \mbox{GeV}/\mbox{fm}$, dash-dotted lines) collisions for
  central rapidity $|\eta| \leq0.5$ at $\sqrt{s_{NN}}=200$ GeV. From
  left to right we show the $p_T$ distributions for the leading (1) and
  the away hadrons (2, 3), which come from partons produced in four different
  momentum bins, from top to bottom: $2-4$ (a), $6-8$ (b), $10-12$ (c) and
  $14-16$ (d) $\mbox{GeV}$.}
\label{fig2}
\end{figure*}

To study $p + p$ collisions, partons are fragmented into hadrons, by
means of KKP fragmentation~\cite{KKP}. To describe medium effects on
the propagation of partons and their hadronization in a nuclear
collision, we include energy loss effects through KKP modified
fragmentation functions~\cite{Zhang}.  In the former case, parton $i$
with momentum $p$ is assigned a random value of momentum fraction $z$,
which allows it to evolve into hadron $H$ with momentum $p^H$. In the
latter, we take into account that the hard scattering can take place
anywhere within the nuclear collision overlap area and that the
produced partons can travel in any direction.  This is implemented by
assigning the hard scattering a random radial position $r$ and a
random direction $\phi$ that the leading parton makes with the outward
radial direction.  These parameters are used to determine the average
number of scatterings $\langle L / \lambda\rangle$, the energy loss
$\Delta E_i$ for each parton and their modified momentum fractions
$z^\prime_i$.  In this way the hadron level events are characterized
by the particle species and their kinematical variables both in $p +
p$ and $A + A$ collisions.  Notice that the KKP fragmentation
functions are valid only for hadron momentum fractions $z$ in the
range $0.1<z<0.8$ which affects the shape of the azimuthal
correlations for the away-side hadrons' relative angular separation
close to $0$ and $\pi$ radians.

For the nuclear environment, since the minimum allowed parton momentum
for hadronization is, according to the model in Ref.~\cite{Zhang},
$p_T=2.4$ GeV, then we have three different types of events
corresponding to the number of partons that are absorbed by the medium
due to energy loss. Notice that this implies that there is a
contribution from $2 \to 3$ partonic processes (where one of the final
state parton is absorbed by the medium) to the di-hadron azimuthal
correlation function.

\section{$p_T$ distributions for $2 \to \{2,3\}$ hadron production}\label{IV}

In order to better characterize parton energy loss, we first build the
$p_T$ distributions for partons produced in given momentum bins. We
then build the hadron $p_T$ distributions that these partons originate
and identify the corresponding momentum bins where these hadrons are
produced.  For the sake of definiteness, we carry out this procedure
for exclusive momentum bins $2-4$, $6-8$, $10-12$ and $14-16 \ \mbox{GeV}$.

Figure~\ref{fig1} shows the parton and hadron $p_T$ distributions at
central rapidity $\left| \eta \right| \leq 0.5$ for $2 \rightarrow 2$
events, both in $p + p$ ($\epsilon_0 = 0)$ and $A + A$
($\epsilon_0=2\ \mbox{GeV}/\mbox{fm}$) collisions at $\sqrt{s_{NN}} = 200$
GeV. On the left (right) column we show the leading (away) hadron
$p_T$ distribution.  The leading hadron has been defined, event by
event, as the hadron with the largest $p_T$, thus for each parton
$p_T$ bin, the away hadron momentum distribution is shifted towards
lower values of $p_T$ when compared to the leading one. The effect is
more pronounced in $A + A$ collisions due to energy loss. Notice also
that momentum conservation at the parton level is reflected by the
fact that the parton momentum distribution is the same for leading and
away side hadrons.

\begin{figure*} 
{\centering
  {\includegraphics[scale=0.3]{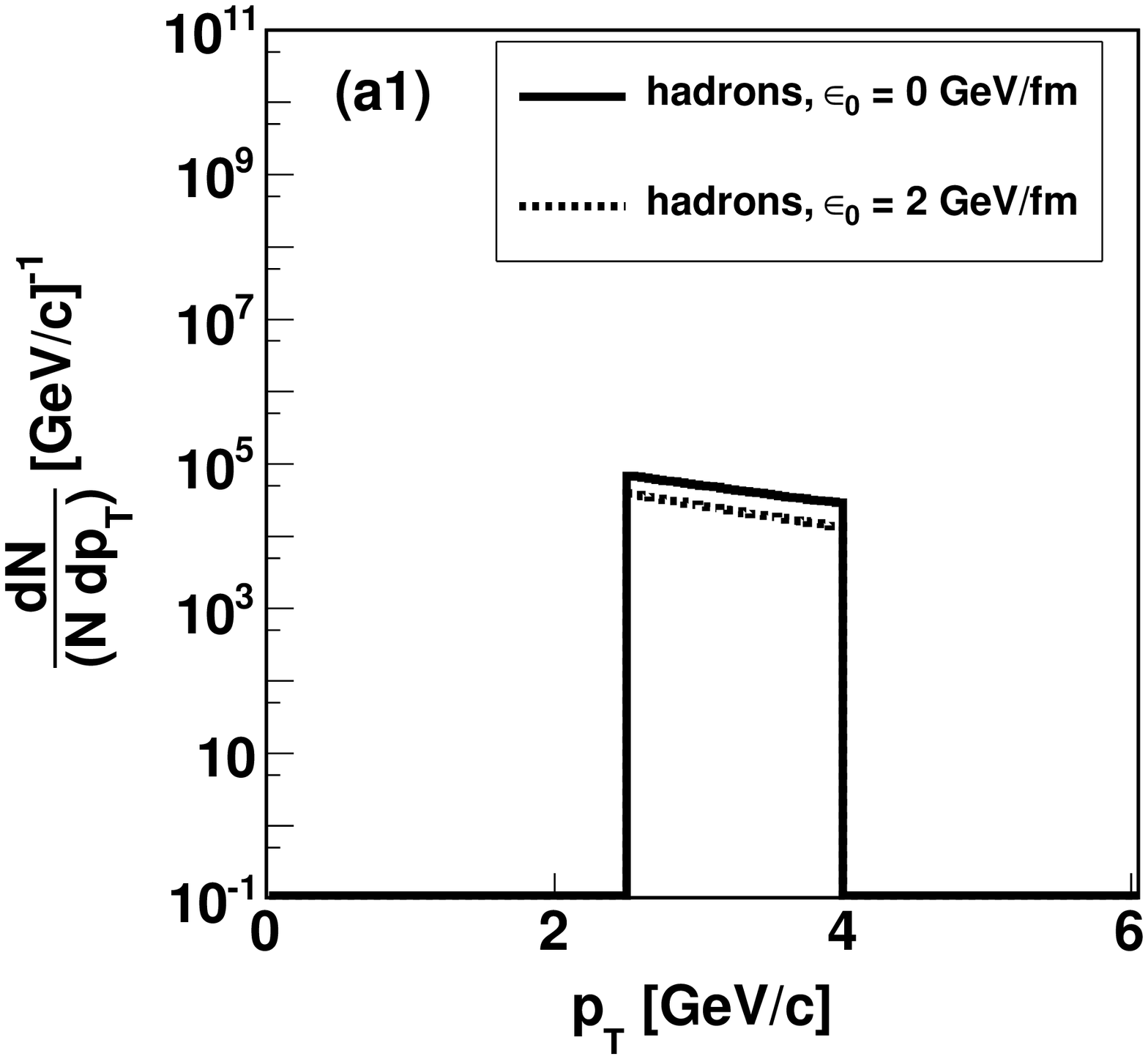}\includegraphics[scale=0.3]{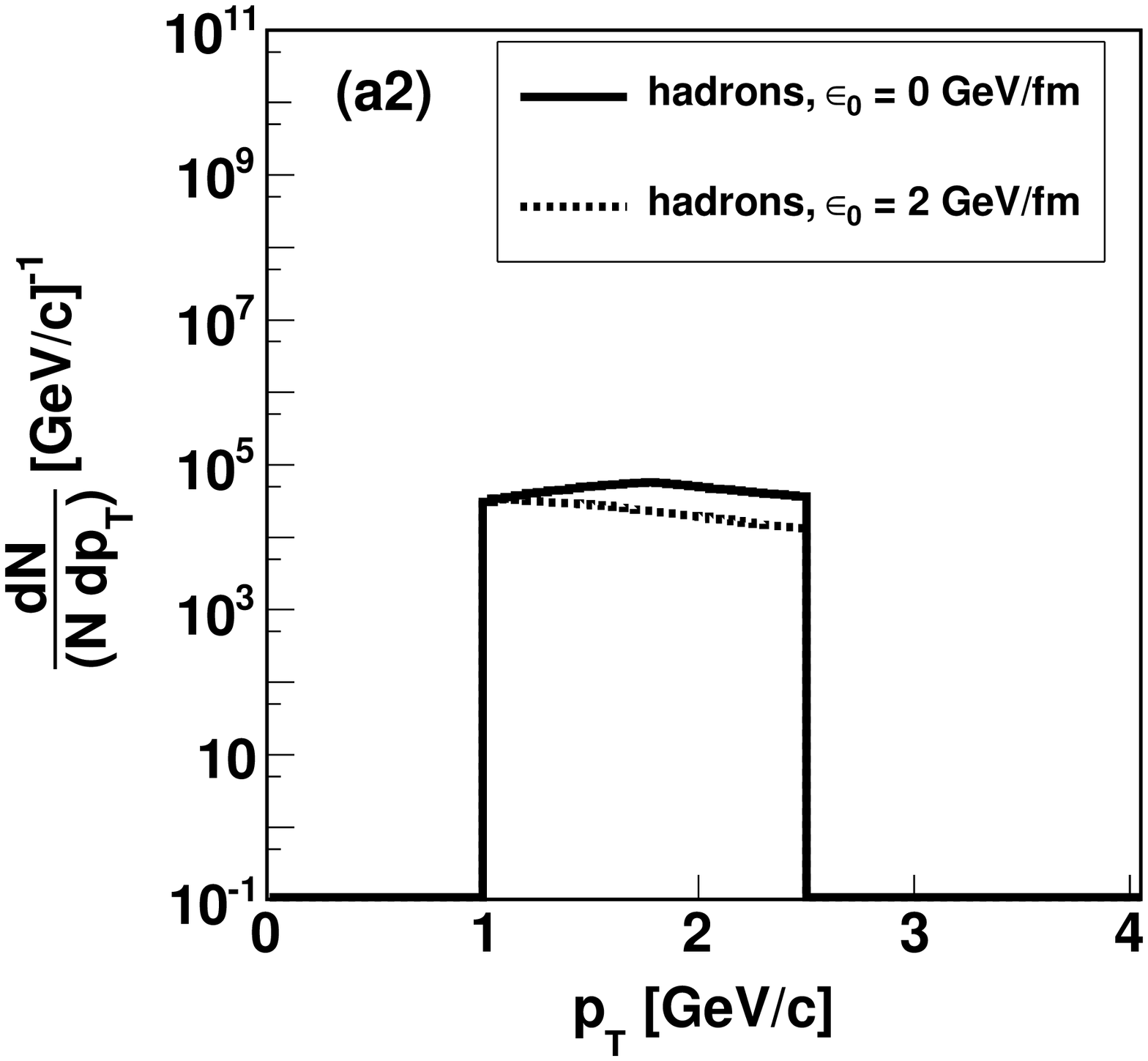}
    \includegraphics[scale=0.3]{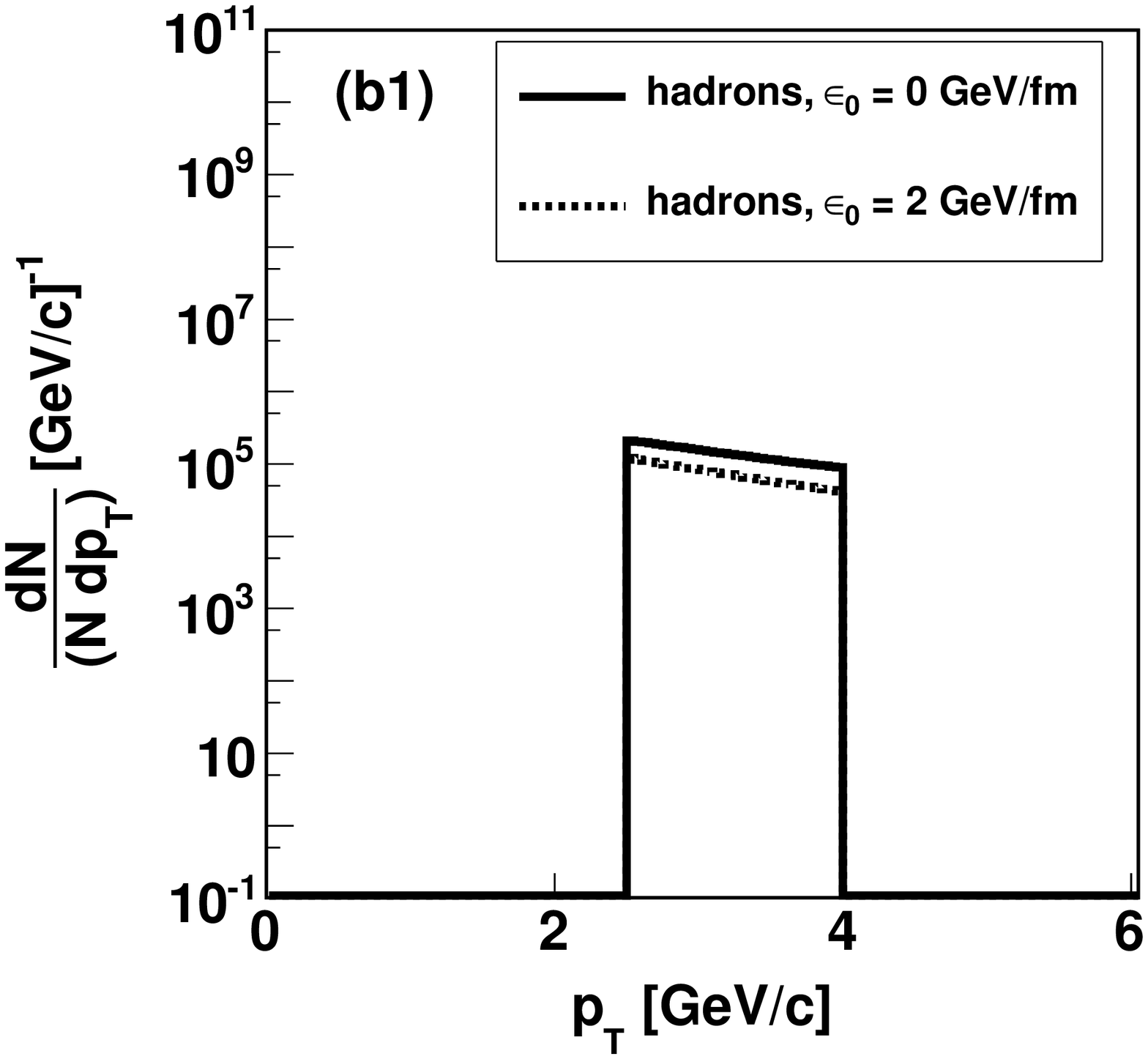}\includegraphics[scale=0.3]{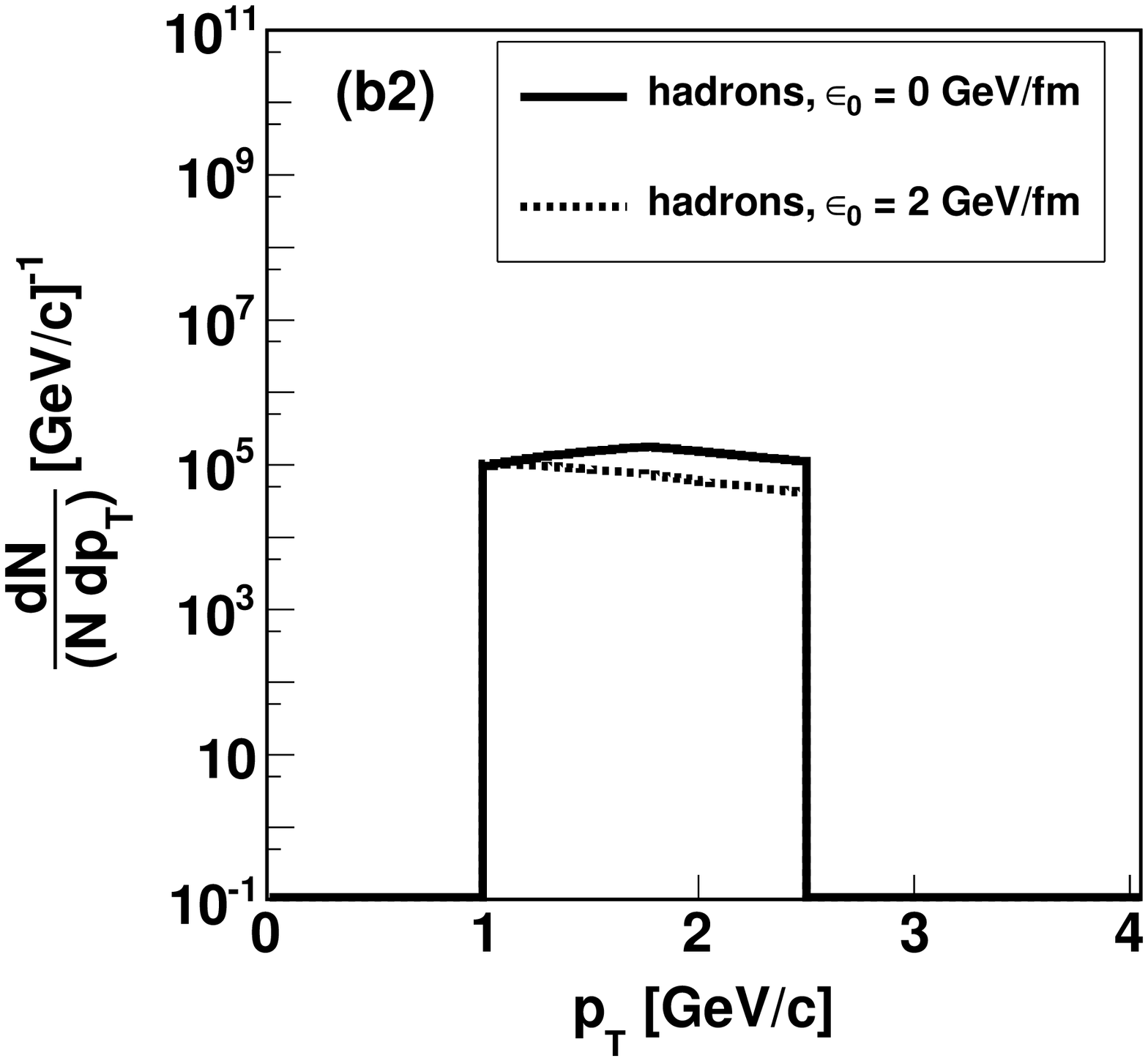}\includegraphics[scale=0.3]{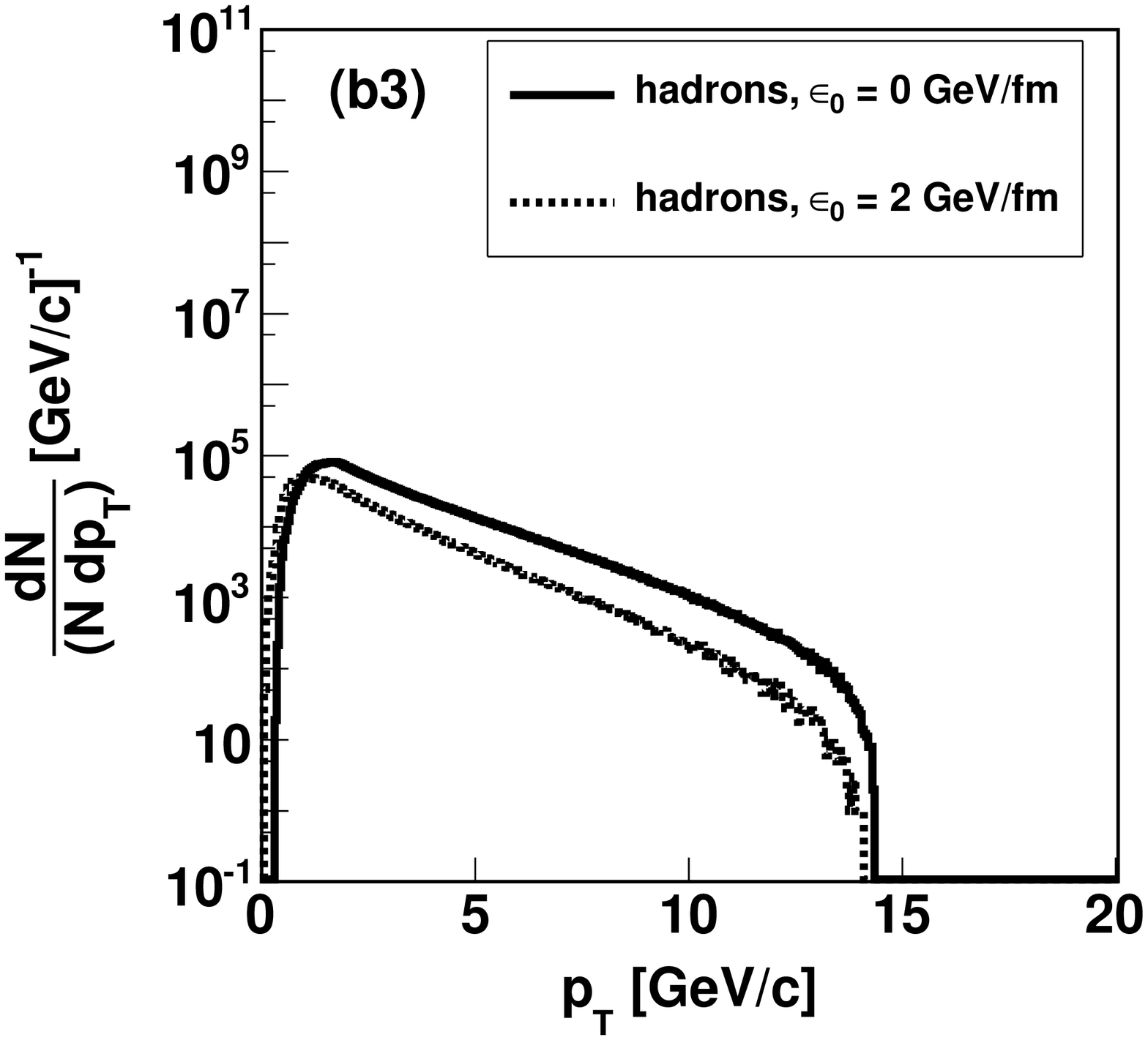}
}}
\caption{Hadron $p_T$ distributions for events that contribute to the
  di-hadron azimuthal correlation. Shown are $2\to 2$ (a) and $2\to 3$
  (b) processes. The solid lines correspond to the case without medium
  and the dotted lines to the case within the medium. In both cases
  the leading hadron is in the momentum bin $2.5\ \mbox{GeV}\leq p_{T}
  \leq 4\ \mbox{GeV}$ (1). For the upper row the away side hadron is
  in the momentum bin $1\ \mbox{GeV}\leq p_{T} \leq 2.5\ \mbox{GeV}$
  whereas in the second row, at least one of the away side hadrons is
  in the momentum bin $1\ \mbox{GeV}\leq p_{T} \leq 2.5\ \mbox{GeV}$
  (2).}
\label{fig3}
\end{figure*}

Figure~\ref{fig2} shows the parton and hadron $p_T$ distributions for
$2 \to 3$ events, both in $p + p$ ($\epsilon_0=0$) and in $A + A$
($\epsilon_0=2\ \mbox{GeV}/\mbox{fm}$) collisions. In the figure, the
left column shows the momentum distribution of the leading hadron
which, as before is the hadron with the largest $p_T$ in the event,
the middle column displays the away side hadron selected as the hadron
with the second largest momentum and the right column shows the away
side hadron with the lowest $p_T$ in the event.  It is also
interesting to note that the bulk of the hadron events with low
momenta (in the range 0.2 GeV $\leq p_T \leq$ 2 GeV), come from parton
events with momenta in the range 2 GeV $\leq p_T \leq$ 10 GeV, whereas
the hadron events with higher momenta (in the range 2 GeV $\leq p_T
\leq$ 4 GeV), come from parton events with momenta in the range 10 GeV
$\leq p_T \leq$ 16 GeV.

We now focus on the characterization of the events that contribute to
the di-hadron azimuthal correlation either from $2\to 2$ or from $2\to
3$ parton processes. Figure~\ref{fig3} shows the hadron $p_T$
distribution calculated under the aforementioned kinematical
conditions, where we use all the parton exclusive momentum bins, $2$
GeV wide, in the range $2-18$ GeV.  The leading hadron is in the bin
$2.5\ \mbox{GeV}\leq p_{T} \leq 4\ \mbox{GeV}$.  In the first row we
show $2\to 2$ events where the away side hadron is in the momentum bin
$1\ \mbox{GeV}\leq p_{T} \leq 2.5\ \mbox{GeV}$ and in the second row
we show $2\to 3$ events where at least one of the away side hadrons is
in the momentum bin $1\ \mbox{GeV}\leq p_{T} \leq 2.5\ \mbox{GeV}$.
We choose these hadron momentum bins to compare with the analysis of
Ref.~\cite{Ajitanand}. Also, the selection of hadron events from
parton $p_T$ bins in specified ranges favors mercedes-like
configurations for hadrons. Notice that as expected, in each case the
distributions without medium are consistently higher than the ones
with medium effects.

\section{Three-hadron azimuthal correlations}\label{V}

We now look at the azimuthal correlations for the away side
considering the contributions from $2 \to 2$ and $2\to 3$ hadron
events.

\begin{figure*} 
{\centering
  {\includegraphics[scale=0.3]{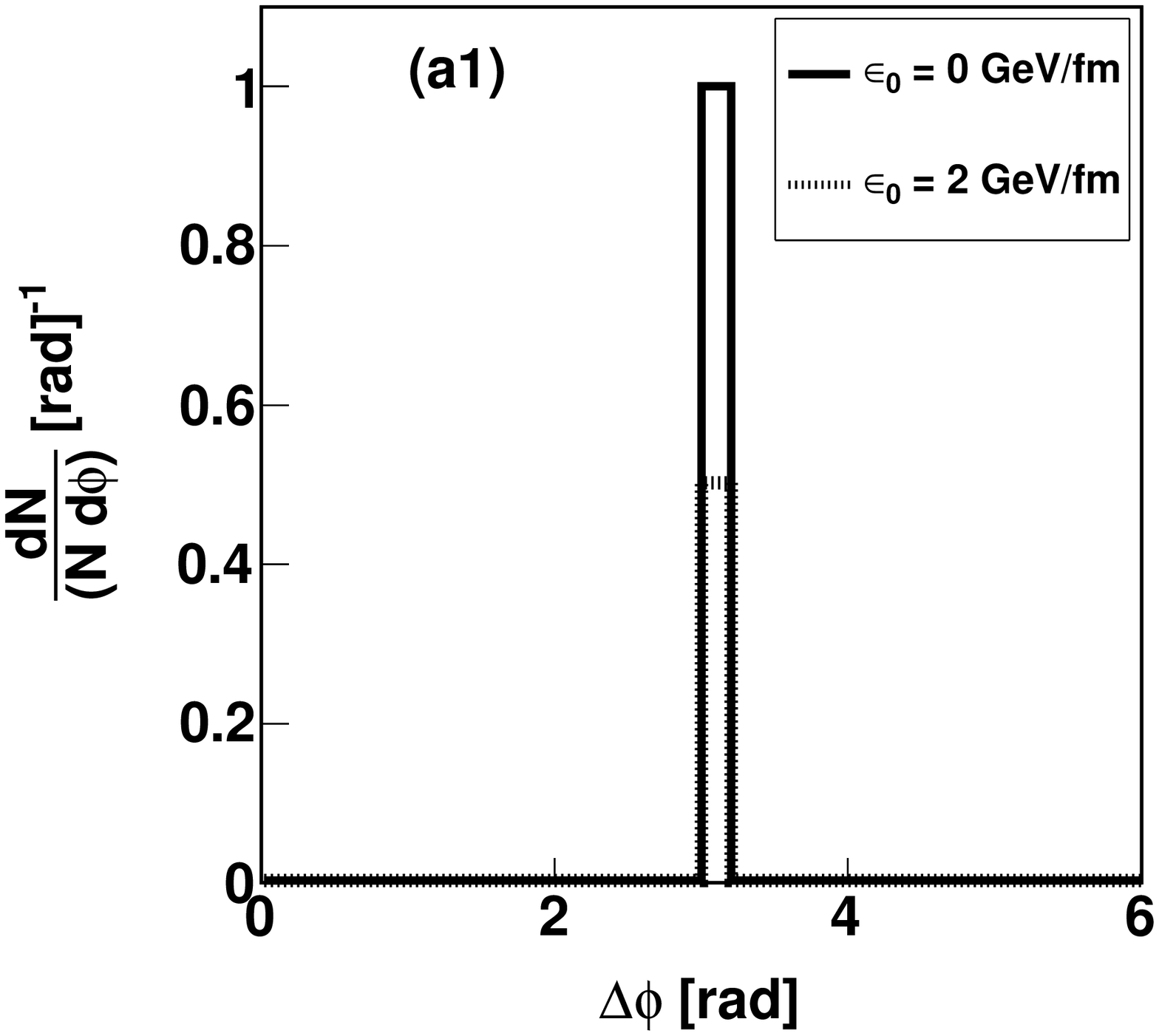}\includegraphics[scale=0.3]{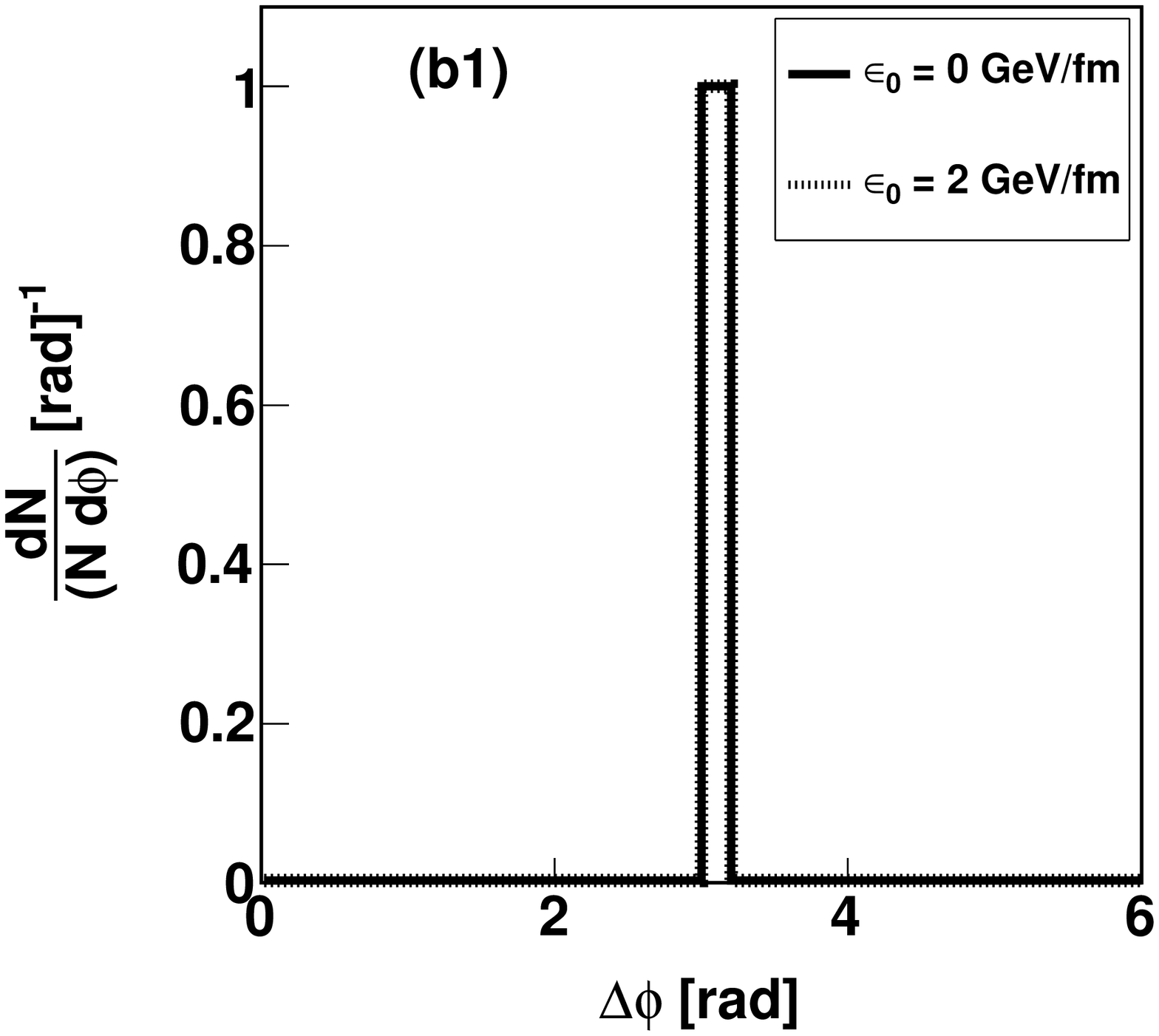}
    \includegraphics[scale=0.3]{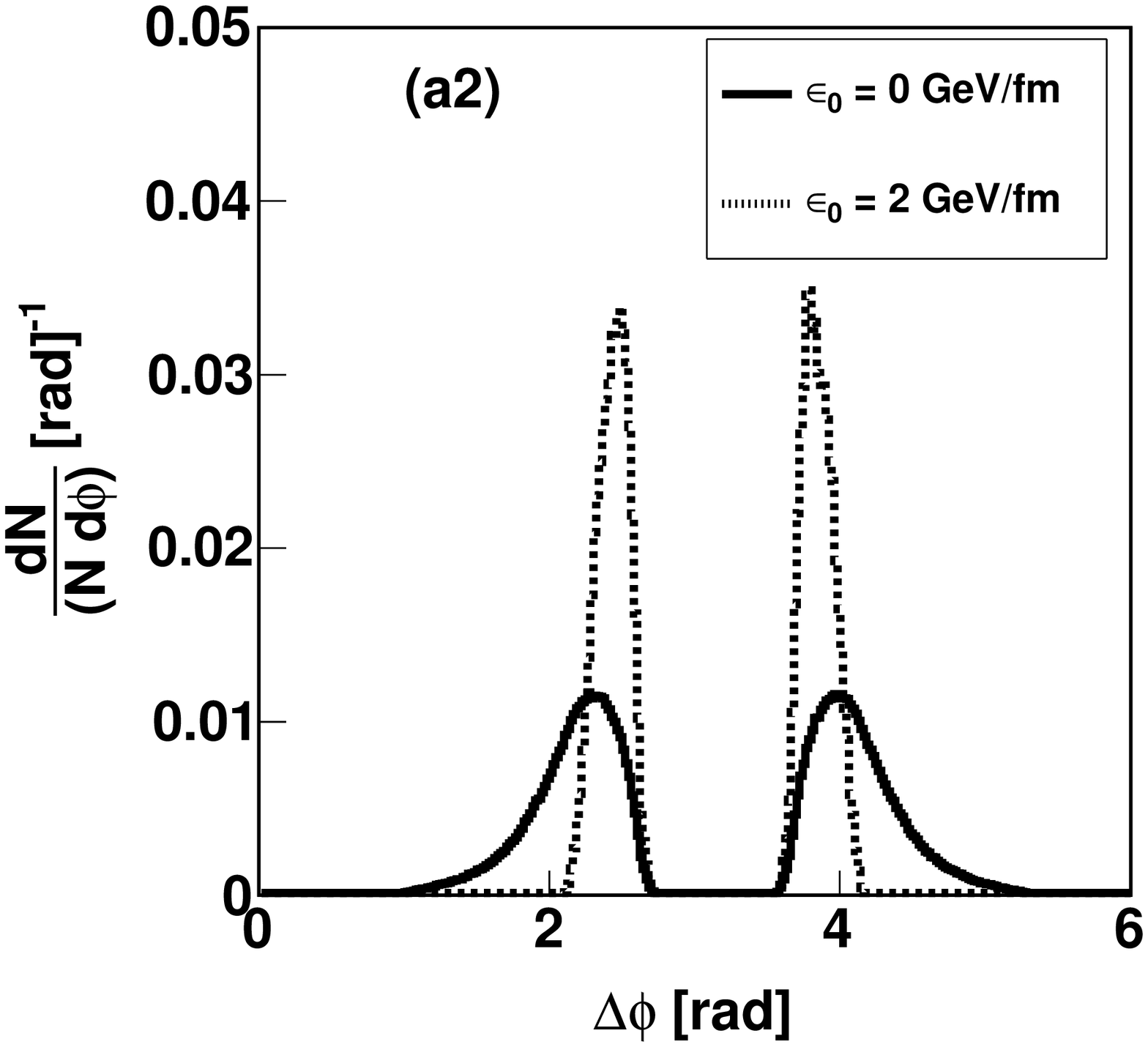}\includegraphics[scale=0.3]{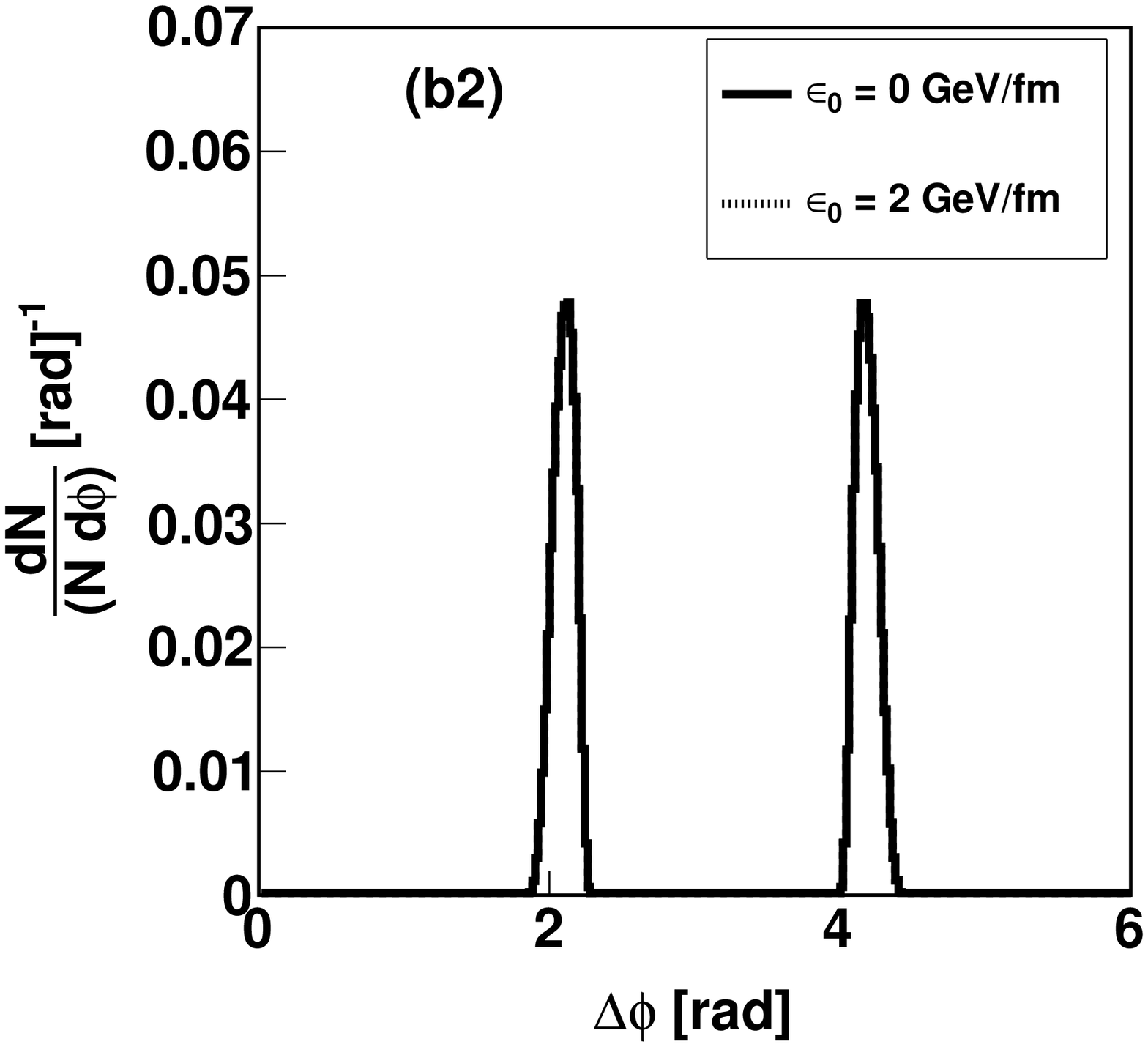}
}}
\caption{Di-hadron azimuthal angular correlation normalized to the
  number of parton events and to the bin size from $2 \to 2$ (1) and
  $2 \to 3$ (2) events in parton $p_T$ bins $2-4$ (a) and
  $10-12\ \mbox{GeV}$ (b) for $p + p$ ($\epsilon_0=0$) and $A + A$
  ($\epsilon_0=2\ \mbox{GeV}/\mbox{fm}$) collisions}
\label{fig4}
\end{figure*}

Figure~\ref{fig4} shows the di-hadron azimuthal angular correlation
normalized to the number of parton events and to the bin size from $2
\to 2$ (upper row) and $2 \to 3$ (lower row) events.  For this figure
we use parton $p_T$ bins $2-4$ and $10-12\ \mbox{GeV}$ for $p + p$
($\epsilon_0=0$) and $A + A$ ($\epsilon_0=2\ \mbox{GeV}/\mbox{fm}$)
collisions at $\sqrt{s_{NN}}=200$ GeV.  Notice the typical peak shape
at $\pi$ radians in the $2\to 2$ case, that in our approach arises due
to collinear fragmentation, whereas in the $2 \to 3$ case the peaks
are located at $2\pi/3$ and $4\pi/3$ radians and have a finite
width. In the $2-4$\ GeV parton $p_T$ bin, there is a marked
difference between the height of the peaks for hadrons produced in the
$p + p$ and $A + A$ cases. This is due to the fact that in the medium,
the model considers a threshold at $p_T^{\mbox{\tiny{thresh}}} = 2.4$
GeV below which the parton is unable to hadronize and this value lies
within the considered $p_T$ bin.  For all parton $p_T$ bins higher
than this one, the correlation of hadrons produced in both cases
coincide, since the threshold value $p_T^{\mbox{\tiny{thresh}}}$ is
outside such bin. This is illustrated with the $10-12$ GeV parton
$p_T$ bin on the right column of Fig.~\ref{fig4}.

\begin{figure} 
{\centering {
\includegraphics[scale=0.4]{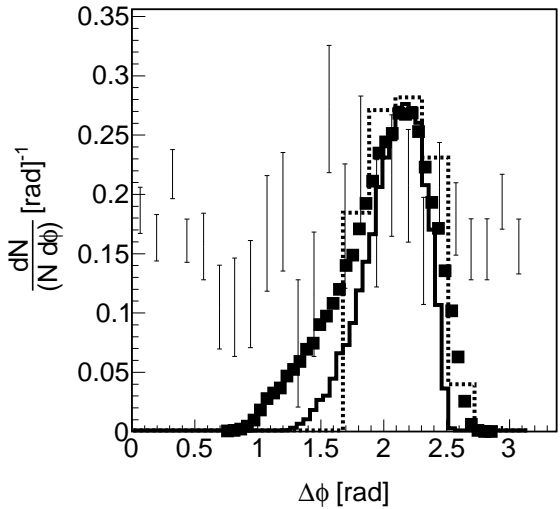} 
}}
\caption{Normalized di-hadron correlation in the nuclear environment
  with $\epsilon_0=2$ GeV/fm as a function of the angular difference
  $\Delta\phi = \theta^H_3 - \theta^H_2$, integrated over angular
  range 1.65 radians $\leq\theta^H_2\leq$ 2.2 radians for a leading
  hadron momentum $2.5$ GeV $\leq p_T\leq 4$ GeV and away side hadron
  momenta $1$ GeV $\leq p_T\leq 2.5$ GeV. The solid line and the
  squares correspond to the Monte Carlo approach with $\epsilon_0=2$
  GeV/fm and $\epsilon_0=0$, respectively, and the dotted line to the
  approach in Refs.~\cite{Ayala} with $\epsilon_0=2$ GeV/fm.  Data are
  from the PHENIX Collaboration~\cite{Ajitanand}. }
\label{fig5}
\end{figure}

Experimental three-particle correlation functions are built from
normalized distributions of a triplet of particles chosen in a single
event. Typically one of these three particles, the leading one, is
chosen in a specified high transverse momentum bin while the two
associated hadrons are chosen in a lower transverse momentum bin. By
fixing the direction of the leading hadron, the analysis is carried
out in terms of the two angles that the away-side hadrons make with
the leading one.  Different correlation scenarios can lead to similar
distributions. To distinguish between scenarios, it is customary to
project the correlation choosing a certain combination of these angles
to remain fixed and an other independent one to vary.

One possibility is to look at the correlation function in terms of the
angular difference $\Delta\phi = \theta^H_3 - \theta^H_2$ of the away
side hadrons for a range of angles of one of the away side hadrons,
say, $\theta^H_2$. Figure~\ref{fig5} shows this correlation in the
proton and nuclear environments with $\sqrt{s_{NN}}=200$ GeV for a
leading hadron momentum $2.5$ GeV $\leq p_T\leq 4$ GeV and away side
hadron momenta $1$ GeV $\leq p_T\leq 2.5$ GeV, integrated over angular
range 1.65 radians $\leq\theta^H_2\leq$ 2.2 radians. Shown are the
normalized histograms for $\epsilon_0=2$ GeV/fm (solid line) and
$\epsilon_0=0$ (squares) obtained from the present Monte Carlo
approach and from the approach in Refs.~\cite{Ayala} with
$\epsilon_0=2$ GeV/fm (dotted line), compared to preliminary data from
the PHENIX Collaboration~\cite{Ajitanand}. Notice that the correlation
peaks at $\Delta \phi=2.2$ radians which is consistent with the
interpretation that for the considered momentum bin, it is more likely
that the two away-side hadrons are produced close to a
\textit{Mercedes-like} configuration. This peak is also present in the
data sample. Also, notice that the Monte Carlo approach histograms
have a more defined peak than the approach where we regulated the
divergences with angular cuts. The difference between the Monte Carlo
result and the data away from $\Delta \phi=2.2$ radians can be
understood by recalling that we have implemented collinear
fragmentation. To be able to populate the regions for $ \Delta\phi
\lesssim 1$\ radians and $ \Delta\phi \gtrsim 2.5$ radians, we would
need to consider the contribution of the leading hadron and go beyond
collinear fragmentation.

\section{Summary and conclusions}\label{VI}

In this work we have presented a Monte Carlo approach to compute
di-hadron correlation functions including contributions from $2\to 2$
and $2\to 3$ parton processes in $p + p$ and $A + A$ collisions. The
production of parton events in $p + p$ is implemented using MadGraph5
where hadron events are obtained by evolving the partons into hadrons
with collinear fragmentation by means of KKP fragmentation.  For the
nuclear environment we follow a modified fragmentation function
approach whereby the hard scattering is assigned a random radial
position $r$ and a random direction $\phi$ that the leading parton
makes with the outward radial direction.  These parameters are used to
determine the average number of scatterings, the energy loss for each
parton and their modified momentum fractions. We use the Monte Carlo
approach to overcome the limitations set by collinear and soft
divergences that otherwise have to be handled by means of angular
cuts. We have studied the hadron $p_T$ ranges that are populated when
partons fragment. To be fully quantitative, one requires to generate
Monte Carlo samples covering a wide range of parton $p_T$, both for $2
\to 2$ and $2 \to 3$ processes. Our approach on the other hand, based
on populating hadron $p_T$ bins from given parton $p_T$ bins although
reducing the need of a large Monte Carlo sample still has limitations,
in particular, to assess quantitatively the relative contributions
between $2\to 2$ and $2\to 3$ processes. Work to overcome this
limitation is underway and will be reported elsewhere.

We conclude that since the energy loss model considers a threshold
parton momentum $p_T^{\mbox{\tiny{thresh}}} =2.4$ GeV below which
partons cannot fragment, there is a need to study the way those
partons interact with the bulk medium and that do not hadronize
through fragmentation. These partons are rather abundant, since the
low momentum part of their spectrum is more populated, and thus their
effect must be considered. When these partons are produced in $2\to 3$
processes, due to momentum conservation they still come out from the
hard scattering mainly with the characteristic mercedes-like
configuration. Their interaction with the bulk partons should produce
a transfer of momentum along these directions that will deform the
medium producing showers of low momentum hadrons bearing the same
distinctive angular distance of the original partons. This avenue is
currently being pursued and the results will also be reported
elsewhere.

\section*{Acknowledgments}

Support for this work has been received in part from DGAPA-UNAM under
grant number PAPIIT-IN103811, CONACyT-M\'exico under grant number
128534, {\it Programa de Intercambio UNAM-UNISON}, the DOE Office of
Nuclear Physics through Grant No.\ DE-FG02-09ER41620, the ``Lab
Directed Research and Development'' grant LDRD~10-043 (Brookhaven
National Laboratory), and by The City University of New York through
the PSC-CUNY Research Award Program, grant 63404-0042. J.J-M. would
like to thank A. Dumitru for discussions on this topic.

\end{document}